\pdfoutput=1
\documentclass[aps,twocolumn,prd,superscriptaddress,showpacs,preprintnumbers,amsmath,amssymb,nofootinbib]{revtex4-1}

\usepackage{amsmath}
\usepackage{amsfonts}
\usepackage{amssymb}
\usepackage{bbm}
\usepackage{dsfont}
\usepackage{mathtools}

\usepackage{graphicx}
\usepackage{slashed}
\usepackage{color}
\usepackage{hyperref}

\newcommand{\imag}{\text{i}}
\newcommand{\skipthis}[1]{}

\newcommand{\Tr}{{\text{Tr}}}

\usepackage[caption=false]{subfig} 
\usepackage[T1]{fontenc}
\usepackage[latin9]{inputenc}
\usepackage{amsmath,amssymb,amsthm,amsfonts}
\usepackage{ mathrsfs }
\usepackage{slashed}
\usepackage{graphicx}
\usepackage{color,rotating}
\usepackage{dsfont}
\usepackage{setspace}
\usepackage{verbatim}
\usepackage{fancyhdr}

\usepackage{MnSymbol}
\usepackage[separate-uncertainty=true]{siunitx}
\usepackage{booktabs}

\usepackage{hyperref}

\def\s0#1#2{\mbox{\small{$ \frac{#1}{#2} $}}}
\def\0#1#2{\frac{#1}{#2}}

\graphicspath{{./figures/}}

\newcommand{\bea}{\begin{eqnarray}}
\newcommand{\eea}{\end{eqnarray}}

\definecolor{darkgreen}{rgb}{0,0.6,0}
\definecolor{gray}{rgb}{.7,.7,.7}

\def\eq#1{(\ref{#1})}
\def\Eq#1{Eq.~(\ref{#1})}
\newcommand {\apgt} {\ {\raise-.5ex\hbox{$\buildrel>\over\sim$}}\ }
\newcommand {\aplt} {\ {\raise-.5ex\hbox{$\buildrel<\over\sim$}}\ }

\def\s0#1#2{\mbox{\small{$ \frac{#1}{#2} $}}}
\def\0#1#2{\frac{#1}{#2}}

\newcommand{\be}{\begin{eqnarray}}
\newcommand{\ee}{\end{eqnarray}}

\newcommand{\del}{\partial}

\newcommand{\fineq}[1]{\;{#1}}							
\renewcommand{\del}[1][]{\partial_{#1}}					
\newcommand{\dd}{\text{d}}							
\newcommand{\GammaTwo}{\Gamma^{(2)}}				
\newcommand{\ie}{i.e.~}
\newcommand{\eg}{e.g.~}

\newcommand{\exppt}{\rho_0}

\newcommand{\symhspace}[2]{\hspace{#1}#2\hspace{#1}}	

\DeclarePairedDelimiter\absVal{\vert}{\vert}

\DeclarePairedDelimiter\chevron{\langle}{\rangle}  

\begin{document}

\title{Towards quantitative precision in the chiral crossover: masses and fluctuation scales}
\author{Alexander J. Helmboldt}
\affiliation{Institut f\"ur Theoretische
  Physik, Universit\"at Heidelberg, Philosophenweg 16, 69120
  Heidelberg, Germany} 

 \author{Jan M. Pawlowski} 
 \affiliation{Institut f\"ur Theoretische
  Physik, Universit\"at Heidelberg, Philosophenweg 16, 69120
  Heidelberg, Germany} 
\affiliation{ExtreMe Matter Institute EMMI, GSI, Planckstr. 1,
  64291 Darmstadt, Germany}

\author{Nils Strodthoff}
\affiliation{Institut f\"ur Theoretische
  Physik, Universit\"at Heidelberg, Philosophenweg 16, 69120
  Heidelberg, Germany}

\begin{abstract}
  We investigate the relation between the physical pion pole and
  screening masses and the mesonic fluctuation scale in low-energy QCD,
  which relates to the curvature of the mesonic effective
  potential. This relation is important for the correct relative
  weight of quantum, thermal and density fluctuations. Hence, it governs
  the location of phase boundaries as well as the phase structure of
  QCD.  The identification of the correct physics scales is also
  primarily important for the correct adjustment of the parameters of
  effective models for low-energy QCD. It is shown that subject to an
  appropriate definition of the latter, all these scales agree at
  vanishing temperature, while they deviate from each other at finite
  temperature.

\end{abstract}
\maketitle

\section{Introduction}

In the past decade rapid progress has been made in our theoretical
understanding of the QCD phase structure, both with continuum methods
and with the lattice. By now, functional continuum approaches
to QCD allow us to discuss the strongly correlated low-energy sector
within a first-principle approach. Then the couplings to be fixed are
simply the fundamental parameters of QCD: the strong coupling
$\alpha_s$ and the current quark masses $m_{\rm current}$, see
\cite{Braun:2009gm,Pawlowski:2010ht,Haas:2013qwp,Herbst:2013ufa}. In
principle, this approach also allows to pin down the coupling
constants in low-energy effective models of QCD as functions of
$\alpha_s$ and $m_{\rm current}$. These models are usually defined at
a (UV) momentum scale $\Lambda_\text{UV}$ of about \SI{1}{GeV} in terms of an
effective Lagrangian with a set of coupling parameters $\vec
\lambda$. As low-energy couplings of QCD they can be deduced uniquely
from QCD as $\vec \lambda(\alpha_s,m_{\rm current})$. Such an approach
is completed by determining the set of all relevant low-energy
coupling parameters (at the UV scale $\Lambda_\text{UV}$) which may have an
impact on the infrared physics at hand.  In summary, this set-up
anchors low-energy models within first-principle QCD and their
independent couplings are only those of QCD.  As a consequence, not
only qualitative but also quantitative physics questions of the
strongly correlated low-energy sector of QCD become accessible.

As described above, it is of chief importance in this set-up to pin
down the relevant couplings in low-energy models. Moreover, the
analysis of low-energy quantum, thermal and density fluctuations in
these models have to be brought to a quantitative level. The current
work does a further significant step in this direction in the context
of the quark-meson (QM) model. We study the frequency and momentum
dependence of two-point correlation functions, which is interesting for
several reasons. Most importantly, it gives us direct access to the
relevant question of physical observables such as pole masses and
decay constants, which are so far only indirectly accessible in the
Euclidean approach.  Within this context it also allows us to
determine and clarify the relations between the physical observables
determined at the poles, e.g.\ at $p^2=-m_\pi^2$ and the low-energy
parameters of the models at $p^2=0$.  In particular, we determine the
relations between pole masses corresponding to propagator poles at
$p_0^2=-m_{\rm pol}^2$, screening masses corresponding to poles at
$\vec p^2=-m_{\rm scr}^2$ and curvature masses $m_{\rm cur}^2$
evaluated at $p^2=0$. Only the latter are directly accessible within
Euclidean approaches and are the mass parameters in the effective
action.

So far, the above relation and its convergence with a given
approximation scheme has not been studied. However, this is chiefly
important for the relative weight of quantum, thermal and density
fluctuations: The characteristic scale of quantum fluctuations is the
curvature mass $m_{\rm cur}$. Below this mass scale the propagation of
quantum fluctuations is suppressed. In turn, the characteristic scale
of density fluctuations is the pole mass $m_{\rm pol}$ (of modes with
non-vanishing quark number). At finite temperature these scales have
some temperature dependence. Thus, the correct identification of
$m_{\rm pol},\,m_{\rm scr}$ and $m_{\rm cur}$ is an important issue in
particular for quantitative approaches towards QCD at finite
temperature and density, where potential mismatches can lead to large
systematic errors.

We discuss this issue at the example of the derivative expansion,
which is an expansion in momenta over mass-scale, $p^2/m^2$. The
derivative expansion is the most popular expansion scheme used in
low-energy effective models. In most applications the local potential
approximation (LPA) is used, where one employs classical
propagators. In contrast, the computation in the present work involves
fully momentum-dependent propagators. This advanced approximation can
be used to resolve apparent inconsistencies reported in the literature
within the LPA, see
\cite{Strodthoff:2011tz,Kamikado:2012bt,Svanes:2010we}.  Finally, we
also evaluate the quality of the LPA$'$ scheme, which includes
momentum-independent wavefunction renormalisation factors, in
comparison to the full calculation.

The strength of our computational approach lies in the fact, that it
provides a stable numerical iteration procedure with only little
numerical overhead compared to the momentum-independent
calculation. From the technical point of view, the method is
applicable to a wide range of possible theories. Furthermore, it can
be extended to complex external momenta along the lines of
\cite{Strodthoff:2011tz,Floerchinger:2011sc,Kamikado:2012bt,Tripolt:2013jra,Tripolt:2014wra}
hence providing direct access to spectral functions in an Euclidean
framework without the need for analytical continuation of given
Euclidean data.

The article is organised as follows. In Sect.~\ref{sec:formalism} we
discuss the embedding of low-energy effective models, and specify the
effective action of the quark-meson model. We also elaborate on the
different mass definitions and their physics content, as well as
providing a brief introduction to the functional renormalisation group (FRG)
and the computational set-up. In Sect.~\ref{sec:results} we present
the results of mass calculations at vanishing and finite
temperature. We discuss the implications for relative fluctuation
scales in particular in view of the chiral phase
boundary. Furthermore, we compare LPA, LPA$'$ and the calculation with
fully momentum-dependent mesonic propagators in terms of quantitative
accuracy. The latter is henceforth referred to as full calculation.

\section{Low-energy QCD and fluctuations}
\label{sec:formalism}

\subsection{Low-energy effective models}
\label{sec:effectiveaction}

We aim at describing the low-energy sector of two-flavor QCD within a
quark-meson model \cite{Ellwanger:1994wy,Jungnickel:1995fp, Berges:1997eu,Schaefer:2004en}. 
As already mentioned in the introduction,
low-energy effective models can be firmly anchored in first-principle
QCD, see
\cite{Braun:2009gm,Pawlowski:2010ht,Haas:2013qwp,Herbst:2013ufa}.  The
key idea behind this embedding in full QCD is the functional RG
approach to QCD with dynamical hadronisation,
\cite{Gies:2001nw,Gies:2002hq,Pawlowski:2005xe,Floerchinger:2009uf}.
There, the flow is initiated at a large momentum scale $\Lambda_{\rm
  UV}\gg \Lambda_{\rm QCD}$, where it starts with the effective action
of perturbative QCD with dynamical quarks and gluons. Then, by
lowering the momentum scale within this first-principle QCD framework,
the hadronic degrees of freedom get dynamical at the hadronisation
scale, while the quark and gluon degrees of freedom decouple. This is
most simply seen in the Landau gauge, where the gluon propagator is
infrared gapped, the gapping being directly related to the QCD mass
gap, see e.g. \cite{Fischer:2008uz}. Accordingly, the gluons can be
integrated out first, leading to an effective theory with quarks and
hadronic degrees of freedom in a gluonic background potential at a
momentum scale $\Lambda_\text{UV}\approx$ \SI{1}{GeV}, such as Polyakov loop
enhanced low-energy models. First results within such a QCD-enhanced
model approach have been presented in
\cite{Herbst:2013ufa,Haas:2013qwp}. This setting entails, that
first-principle QCD flows can be employed to provide initial
parameters and further glue input, such as background potentials, for
model calculations, thereby systematically removing ambiguities in
these approaches.

The QCD computation relies only on two input parameters, the strong
coupling $\alpha_s$ and the current quark mass $m_\text{current}$, and
allows to consistently include quantum and thermal fluctuations, where
hadronic correlations are captured via the dynamical hadronisation. In
this way model computations can profit directly from a systematic
improvement in predictive power. Conversely, the quantitative advances
in model calculations, such as put forward in this work, can be easily
carried over to full QCD computation. This allows to systematically
improve the approximation of the first-principle QCD flows in the
low-energy regime. 

In this work we study a quark-meson model taking into account the
momentum dependence of mesonic two-point functions. At vanishing
temperature, the model in the presence of low-energy quantum
fluctuations is approximated by an effective action of the form
\begin{equation}
\begin{split} 
\Gamma=&\int_x \Bigl\{ \bar\psi (\slashed \partial+h (\sigma+\imag 
\gamma^5\vec \tau \cdot \vec \pi))\psi\\
&+\tfrac{1}{2}Z\partial_\mu \phi_i \partial_\mu 
\phi_i+\tfrac{1}{8}Y\partial_{\mu}\rho\partial_{\mu}\rho
+ U(\rho) \Bigr\}\,.
\label{eq:effectiveaction}
\end{split}
\end{equation}
Here $\rho=\sigma^2+\vec\pi^2$, and we include momentum- and field-dependent 
bosonic wavefunction renormalisation factors $Z=Z(p^2;\rho)$
and $Y=Y(p^2;\rho)$. The bosonic sector corresponds to a fluctuating
$O(N)$ model \cite{Wetterich:1991be,Berges:2000ew}, while the fermionic sector is classical, anticipating
the decoupling of quark fluctuations at low energies and
temperatures. In the following, we expand $Z$ and $Y$ about a field
value $\exppt$, restricting ourselves to the zeroth order terms, i.e.\
$Z=Z(p^2;\exppt)$ and $Y=Y(p^2;\exppt)$. However, we take into account
the full momentum dependence of $Z,Y$, as well as computing a full
effective potential $U(\rho)$. The quantitative accuracy of a low
order in the field-expansion of $Z$ has been tested in
\cite{Pawlowski:2014zaa}, which is in this work extended by the
inclusion of $Y$. The mesonic two-point functions, evaluated at
a constant field configuration $\phi_i=(\sqrt{\rho},\vec 0)_i$, read
\begin{equation}
\begin{split}
  \Gamma^{(2)}_{\pi_{i}\pi_{j}}&=\delta_{ij}Z_{\pi} p^{2}
  +2U'(\rho)\delta_{ij}\,,\\
  \Gamma^{(2)}_{\sigma\sigma}&=Z_{\sigma} p^{2}
  +2U'(\rho)+4U''(\rho)\rho\,,
\end{split}
\end{equation}
where $Z_{\pi}=Z(p^2;\exppt)$ and 
$Z_{\sigma}=Z(p^2;\exppt)+Y(p^2;\exppt)\rho$.

\subsection{Pole-, screening- and curvature masses}
\label{sec:massdefinitions}

Particle masses can be extracted directly from the fully
momentum-dependent propagators. In this section, we review different
mass definitions based on the renormalised inverse two-point function
$\bar\Gamma^{(2)}(p_0,\vec p^2)=\Gamma^{(2)}(p_0,\vec p^2)/\bar
Z$. Here, the momentum-independent wavefunction renormalisation $\bar
Z$ relates the bare field $\phi$ to the renormalised field
\begin{equation}\label{eq:WFR}
\bar \phi=\bar Z^\frac{1}{2}\phi\,.
\end{equation}
At vanishing temperature the wavefunction renormalisation $\bar Z$ is
directly related to $Z=Z_\pi$ in \eq{eq:effectiveaction} evaluated at some 
fixed external momentum. In turn, at finite
temperature the heat bath singles out a rest frame, and the wave
function renormalisation $Z$ splits into one component parallel,
$Z_\parallel$, and one perpendicular, $Z_\perp$, to the heat
bath. Accordingly, we parameterise the inverse propagator (at
$\rho=\rho_0$) as
\begin{equation}
\label{eq:AnsatzGamma2}
\Gamma^{(2)}(p_0,\vec p^2)=Z_\parallel(p_0,\vec p^2)p_0^2
+Z_\perp(p_0,\vec p^2)\vec p^2+m^2 \,,
\end{equation}
with $Z_\parallel, Z_\perp$ being finite for vanishing momentum and/or frequency.  
While the decomposition in \eq{eq:AnsatzGamma2} into
$Z_\parallel(p_0,\vec p^2)$ and $Z_\perp(p_0,\vec p^2)$ is not unique for general momenta, it is for
vanishing $|\vec p|$ or $p_0$. Hence, we can define momentum-independent
wavefunction renormalisation factors parallel and perpendicular to the
heat bath within an evaluation at $p_0=0$ or
$\vec{p}=0$, respectively, i.e.\
\begin{equation}
\label{eq:wffactors}
\begin{split}
Z_\parallel&=\lim_{p_0\to 0}\frac{\Delta\Gamma^{(2)}(p_0,0)}{p_0^2} \,,\\[2ex]
Z_\perp&=\lim_{|\vec p|\to 0}\frac{\Delta\Gamma^{(2)}(0,\vec p^2)}{\vec p^2} \,,
\end{split}
\end{equation}
where $\Delta \Gamma^{(2)}(p_0,\vec p^2)\equiv \Gamma^{(2)}(p_0,\vec
p^2)- \Gamma^{(2)}(0,0)$. Then a standard choice at finite temperature
for the renormalisation of the field is $\bar Z=Z_\perp$.  This
definition interpolates between the $O(4)$-symmetric definition at
$T=0$ with $\bar Z= Z_\perp =Z_\parallel$ to that of the dimensionally
reduced theory for $T\to \infty$, where the propagation is
perpendicular to the heat bath. This choice is based on the mass scale
of spatial quantum fluctuations. The ratio $Z_\|/Z_\bot$ gives the
relative weight of the temporal and spatial fluctuations, and hence is
susceptible to the difference of the scale of thermal and quantum
fluctuations.

Now we define pole ($m_{\text{pol}}$), screening ($m_{\text{scr}}$),
and curvature ($m_{\text{cur}}$) masses via
\begin{equation}
\begin{split}
\bar\Gamma^{(2)}(\imag m_{\text{pol}},0)&=0\,,\\[2ex]
\bar\Gamma^{(2)}(0,|\vec p|^2=- m_{\text{scr}}^2)&=0\,,\\[2ex]
\bar\Gamma^{(2)}(0,0)&=m_{\text{cur}}^2\,,
\end{split}
\label{eq:masses}
\end{equation}
assuming propagator poles at real $p^2$. The masses $m_{\rm pol}^2$,
$m_{\rm scr}^2$ are the solutions to \eq{eq:masses} with the minimal
distance to $p^2=0$, in general they have a minimal distance to the
Euclidean frequency axis. Accordingly, the pole and screening masses
are respective inverse temporal and spatial screening lengths. For
minimal distance poles at $\pm\imag m_{\text{pol}}$ and $\pm \imag
m_{\text{scr}}$, we find an exponential decay of the propagator in
position space,
\begin{equation}
\begin{split}
  T \sum_{p_0} \left[\Gamma^{(2)}(p_0, 0)\right]^{-1}e^{\imag p_0 t}
  &\sim e^{-m_{\text{pol}} |t|}\,,\\[2ex]
  \int d^3 p \left[\Gamma^{(2)}(0,\vec p^2)\right]^{-1}e^{\imag \vec p
    \cdot \vec x}&\sim e^{-m_{\text{scr}} |\vec{x}|}\,,
\end{split}
\end{equation}
for $|t|\to\infty$ and $|\vec{x}|\to\infty$, respectively. At
vanishing temperature $\Gamma^{(2)}$ is a function of the $O(4)$
invariant $p_0^2+\vec p^2$ and hence pole and screening masses agree
by definition, i.e.\ $m_{\text{pol}}=m_{\text{scr}}$. At finite
temperatures, the ratio of pole and screening masses is given by
\begin{equation}
\label{eq:ratiopolescr}
\frac{m_{\text{pol}}^2}{m_{\text{scr}}^2}=\frac{Z_\perp(0,\vec p^2=
  - m_{\text{scr}}^2)}{Z_\parallel(\imag m_{\text{pol}}, 0)}\,.
\end{equation}
Due to the breaking of Euclidean $O(4)$-invariance via the heat bath,
the ratio \eq{eq:ratiopolescr} takes values different from unity. In
general it is also different from
$Z_\parallel/Z_\perp=Z_\parallel(0,\vec 0)/Z_\perp(0,\vec 0)$ which is
accessible in finite-temperature LPA$'$ calculations, e.g.\
\cite{Braun:2009si}. Naturally pole and screening masses take
different values at finite temperature. Moreover, these differences
due to the breaking of Euclidean $O(4)$-invariance extend to the
momentum and frequency dependence and to finite chemical potential.

Note that contrary to pole and screening masses, which are directly
physics observables, the curvature mass is not. This is already
obvious from the fact that it depends on the renormalisation
prescription. For $\bar Z = Z_\perp$ it relates to the screening mass,
but is not identical with the latter.  A more detailed discussion is
presented below in the context of the finite temperature case.

Finally, in the vacuum in relativistic theories one also has the onset
mass $m_\text{ons}$. Its definition exploits the fact that the
critical chemical potential associated to the onset of a condensation
phenomenon is linked to the pole mass $m_{\rm pol}$ of the lightest
resonance with non-vanishing quark or baryon number via an exact
Silver Blaze argument \cite{Cohen:2003kd}. The onset mass
$m_\text{ons}$ coincides with the lightest pole mass in the quark
propagator by a Silver Blaze argument, which can be shown in any
diagrammatic expansion scheme in full propagators such as the
functional renormalisation group approach or a 2PI-expansion. The
chemical potential enters the propagator as an imaginary shift of the
zero momentum component. Hence, there is no dependence of the
frequency integrals on the chemical potential until the chemical
potentials exceeds the closest singularity to the real (Euclidean)
$p_0$-axis of the quark propagator in the complex $p_0$-plane. The
position of this singularity coincides with the pole mass of the
corresponding resonance. This agreement between pole and onset mass
was checked explicitly in \cite{Strodthoff:2011tz} in a numerical calculation 
within the LPA. We emphasise that in such a
diagrammatic approach diagrams contributing to the meson propagators or
higher correlation functions with vanishing quark number are not
directly sensitive to the chemical potential up to the onset mass in
the quark propagator.

It remains to establish a connection between pole and curvature
mass. With our choice $\bar Z=Z_\bot$ we find 
\begin{align}\label{eq:mpolmcur}
  m_{\text{cur}}^2=\frac{Z_\parallel(\imag m_{\text{pol}},0)}{\bar{Z}}
  m_{\text{pol}}^2= \frac{Z_\parallel(\imag
    m_{\text{pol}},0)}{Z_\perp(0,0)}m_{\text{pol}}^2 \,.
\end{align}
In particular, at zero temperature the ratio
$m_{\text{cur}}^2/m_{\text{pol}}^2$ is given by the ratio $Z(-
m^2_\text{pol})/Z(0)$. This entails that pole and curvature masses
still agree approximately at vanishing temperature if the momentum
dependence of $Z(p^2)$ is rather mild for $|p^2|< m_{\rm
  pol}^2$. Assuming a well-behaved analytic continuation for these
momenta this is tightly linked to a mild momentum dependence of
$Z(p^2)$ on Euclidean momenta $p^2\geq 0$.

We emphasise again that the curvature mass is not defined uniquely. In
particular, we may define $\bar Z= Z_\parallel(\imag
m_{\text{pol}},0)$, for which both definitions agree. This is to be
expected from the K{\"a}ll{\'e}n-Lehmann spectral representation which
relies on expanding about the particle pole. 

At zero temperature the equality $m_\text{pol}=m_\text{cur}$ is best 
achieved by parameterising the
inverse propagator as $\Gamma^{(2)}(p^2)=Z(p^2)(p^2+m_\text{pol}^2)$
and by renormalising the fields with $Z(\smash{p^2}=0)$. For
computational convenience, we choose $\bar Z=Z_\bot$ with
\eq{eq:AnsatzGamma2}, \eq{eq:wffactors}. Then the equality is not
guaranteed and hence all statements about the approximate equality of
these masses should be understood as statements about the mild
momentum dependence of the wavefunction renormalisation. As a final
remark, in truncation schemes with momentum-independent wavefunction
renormalisation factors, such as LPA or LPA$'$, pole and curvature
mass naturally agree if one chooses to renormalise with $\bar{Z}=Z$.

\subsection{Flow equations and momentum dependence}
\label{sec:frg}

For the computation of the effective potential $U(\rho)$ and the
momentum- and frequency-dependent two-point functions
$\Gamma^{(2)}(p_0,\vec p^2)$ we use the functional renormalisation group, 
for QCD-related reviews see
\cite{Berges:2000ew,Pawlowski:2005xe,Gies:2006wv,Schaefer:2006sr,Braun:2011pp}. It
is based on the Wilsonian idea of integrating fluctuations momentum
shell by momentum shell. Technically this is achieved by
introducing an IR regulator function $R_k$ which suppresses quantum
fluctuations from momentum modes with momenta smaller than some RG
scale $k$ which is subsequently taken from some large UV scale to
zero. The evolution of the scale-dependent analogue of the effective
action $\Gamma$, the effective average action $\Gamma_k$, is described
by a simple 1-loop equation involving full field-dependent propagators \cite{Wetterich:1992yh},
\begin{equation}
\label{eq:floweq}
\partial_t\Gamma_k[\psi,\phi]=\frac{1}{2}\text{Tr}\,
\frac{1}{\Gamma_k^{(2)}[\psi,\phi]+R_k}\partial_t R_k,
\end{equation}
where $\Gamma_k^{(2)}$ denotes the second functional derivative of
$\Gamma_k$ with respect to the fields and $t =\log k/\Lambda$ with
some reference scale $\Lambda$. The trace $\Tr$ sums over momenta and
frequencies, internal indices as well as over field species including
the standard relative minus sign for the fermionic loop. The
corresponding functional equations can rarely be solved exactly. For
the present task we resort to the ansatz \eq{eq:effectiveaction},
which includes momentum- and frequency-dependent wavefunction
renormalisation factors $Z_k$ and $Y_k$ as well as a full effective
potential $U_k$. Approximations with full momentum and frequency
dependence have been applied since long within the FRG, see e.g.\
~\cite{Ellwanger:1995qf,Bergerhoff:1997cv,Pawlowski:2003hq,%
  Blaizot:2005wd,Fischer:2008uz,Fister:2011uw}, applications in higher
orders of the derivative expansions are found in e.g.\
\cite{Canet:2003qd,Litim:2010tt}. In the present work we suggest a new
iterative procedure for the solution of fully momentum- and field-dependent 
approximations with relatively small numerical costs. 

\begin{figure}[t]
\centering
    \includegraphics[width=0.97\columnwidth]{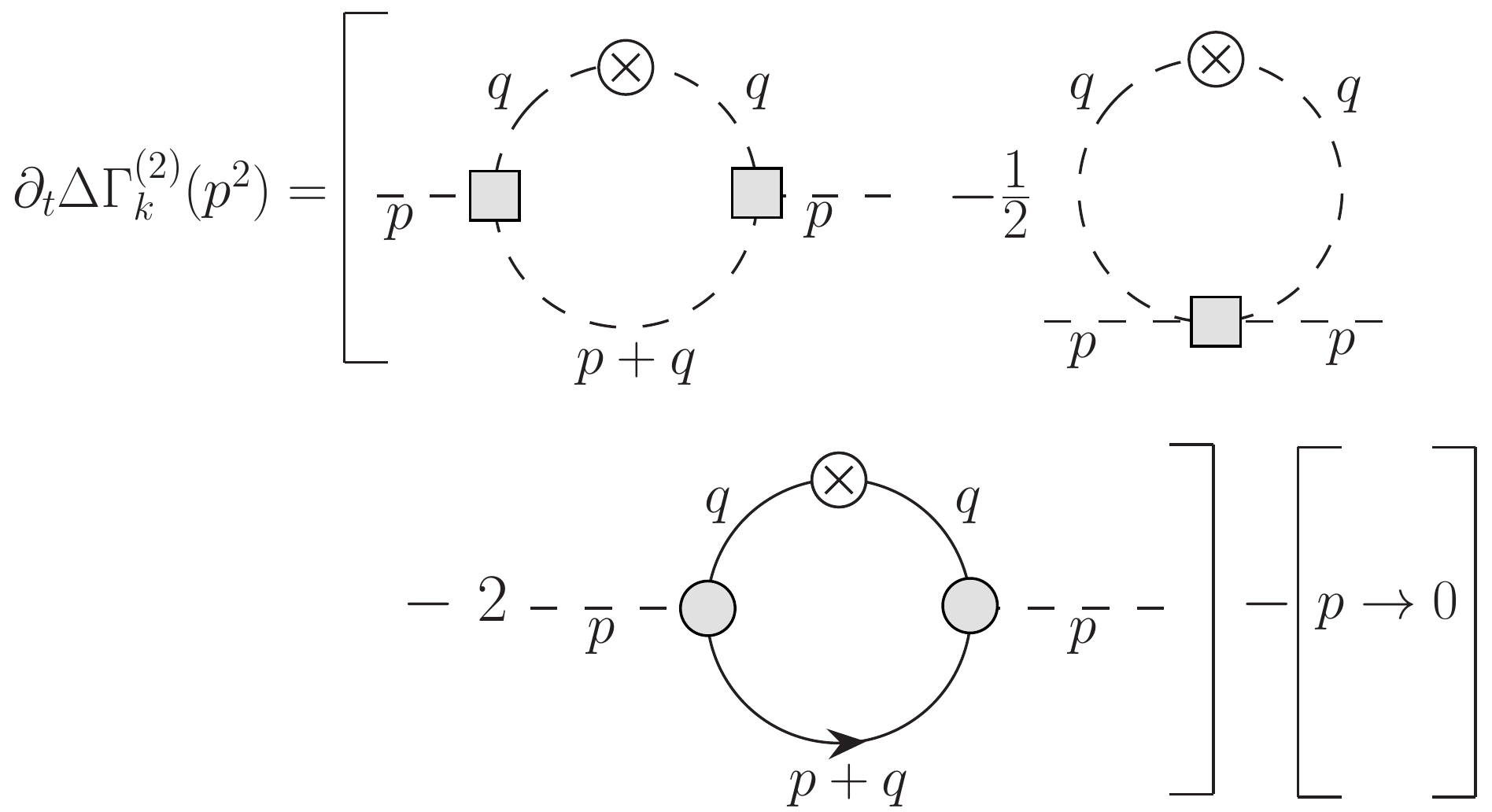}
    \caption{Flow of the momentum-dependent part of the two-point
      function $\Delta \Gamma^{(2)}$.  Dashed (solid) lines denote full
      mesonic (quark) propagators and crossed circles correspond to
      insertions of $\partial_t R_k$ of the respective fields. Tadpole 
	contributions cancel in the present truncation with momentum-independent
mesonic vertices.}
    \label{fig:flow2ptfn}
\end{figure}

Furthermore, although we do not include a genuine running of the
Yukawa coupling as in \cite{Pawlowski:2014zaa}, one either computes at
a fixed bare or renormalised Yukawa coupling. However, in full QCD
flows the renormalised Yukawa coupling stays approximately constant
\cite{Mitter:2014mat}. Hence we employ the latter choice, further
details can be found in App.~\ref{app:convergence}. The expressions
for the inverse propagators, see \eq{eq:AnsatzGamma2}, generalise at
finite $k$ similarly to the effective potential.  The flow equations
for the inverse two-point functions can be obtained from
\eq{eq:floweq} by taking two functional derivatives and are
represented diagrammatically in Fig.~\ref{fig:flow2ptfn}. In general,
these flow equations involve three- and four-point vertices as input,
which are in our case computed from the effective potential, see
App.~\ref{app:flow_equations} for details on the truncation scheme.
\begin{figure}[b]
	\centering
	\includegraphics[width=0.8\columnwidth]{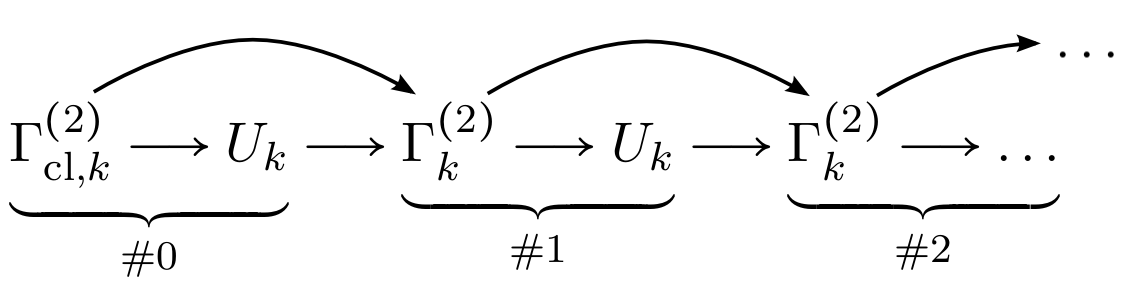}
	\caption[Iteration Scheme]{Illustration of the iteration procedure.}
	\label{fig:iterationScheme}
\end{figure}
In our truncation the flow equation for the effective potential takes
the schematic form
\begin{align}
	\partial_t U_k(\rho) = \mathcal{F}_U \left[ U_k, 
\Delta \GammaTwo_k\right](\rho) \,
	\label{eq:iter:flowEffPot_general}
\end{align}
where
$\Delta\Gamma_k^{(2)}(q^2)=\GammaTwo_k(q^2)-\GammaTwo_k(0)$
at $\rho=\rho_0$. Its flow is given by
\begin{align}
	\partial_t \Delta\GammaTwo_k(p^2) = \mathcal{F}_{G} 
\left[ U_k, \Delta \GammaTwo_k\right](\rho_0,p^2) \,.
	\label{eq:iter:flowGammaTwo_approx}
\end{align}
The explicit form of the flow equations is specified in
App.~\ref{app:flow_equations}. In the following, we expand
\eq{eq:iter:flowEffPot_general} about a field value
$\exppt$. Then \eq{eq:iter:flowEffPot_general} and
\eq{eq:iter:flowGammaTwo_approx} constitute an equation system which
can be solved by an iterative procedure, which is illustrated
pictorially in Fig.~\ref{fig:iterationScheme} and described in App.~\ref{app:iteration}. 
In general, this
iterative method enjoys very good convergence properties. Concerning
the expansion in powers of the field, it has been shown in
\cite{Pawlowski:2014zaa} that an expansion about a fixed bare field
has the best convergence properties. For further details we
refer the reader to App.~\ref{app:convergence}.

\section{Results} \label{sec:results}

In the present section we discuss results within the converged
iterative method introduced above. The approach put forward in the
previous section allows us to discuss the relation between the
different fluctuation scales present in the mesonic sector of QCD:
despite its non-uniqueness the curvature mass relates to the
fluctuation scale of quantum fluctuations. The ratios of
$Z_\perp/Z_\parallel$, and $m_{\rm cur}/m_{\rm scr}$, are a measure
for the relative strength of thermal fluctuations while the pole mass
is the fluctuation scale of density fluctuations.

Additionally, we are interested in the question after the simplest
approximation that already includes all quantitative effects.

\subsection{Masses and fluctuation scales} \label{sec:results:massesandfluctuations}

\begin{table}[b]
	\centering
	\begin{tabular}{cccc}
          \toprule
          \symhspace{1mm}{step} & \symhspace{1mm}{$m_\text{cur}$ [MeV]} & \symhspace{1mm}{$m_\text{pol}$ [MeV]} & \symhspace{1mm}{$\sigma_\text{min}$ [MeV]}\\
          \midrule
          0 & 198.1 & 198.1 &58.2\\
          1 & 135.2 & \num{133+-2}&92.5 \\
          2 & 135.3 & \num{133+-2}&92.8\\
          3 & 135.3 & \num{133+-2}&92.9 \\
          4 & 135.3 & \num{133+-2}&92.8\\
          5 & 135.3 & \num{133+-2}&92.9 \\
          \bottomrule
	\end{tabular}
	\caption{Pion curvature and pole masses and the minimum of the effective potential for different iteration steps at $T=0$ for a physical parameter set at $\Lambda_\text{UV}=\SI{500}{MeV}$. 
          The UV parameters are tuned such that the physical pion mass emerges as the fully converged result. 
          Using the same parameters for a LPA$'$ calculation one obtains $m_\text{pol}=m_\text{cur}=\SI{135.0}{MeV}$.}
	\label{tab:results:massesT0_tune_polLambda500}
\end{table}

At zero temperature, it is sufficient to numerically investigate the
relative size of pole and curvature mass. Screening and pole mass are
equal due to Euclidean $O(4)$ invariance. This holds true for all
cutoff scales and iteration steps, as the regulators used here
preserve $O(4)$ symmetry. As a first non-trivial finding we obtain
rather similar pole and curvature masses, with relative deviations of
less than one percent, see
Tabs.~\ref{tab:results:massesT0_tune_polLambda500}-\ref{tab:results:massesT0_tune_polLambda900}
for calculations at different UV cutoff scales. This result is tightly
linked to a mild momentum dependence of the wavefunction
renormalisation, see the discussion after \eq{eq:mpolmcur}.

The larger the UV cutoff, the more difficult the iteration procedure
gets from the numerical point of view and the slower the convergence
within the iteration procedure gets. This is illustrated in
Tab.~\ref{tab:results:massesT0_tune_polLambda500} and
Tab.~\ref{tab:results:massesT0_tune_polLambda900}, where the
calculation at the smaller cutoff scale converges practically after
the first iteration step, whereas the calculation at the larger cutoff
scale requires more than three iteration steps until approximate
convergence is reached. The reason for the slower convergence has to
be found in the fixed renormalised Yukawa coupling as a calculation
with a fixed bare Yukawa coupling, irrespective of the chosen UV
cutoff scale, shows similarly good convergence properties as our
calculation at the lowest cutoff scale, see
App.~\ref{app:convergence}. In a calculation with a fixed renormalised
coupling, as employed here, the fermionic contribution, as the
dominant contribution to the flow at large cutoff scales, is no longer
decoupled from the bosonic parts of the model but dependent on the
wavefunction renormalisation $Z$. In particular, there is a relevant
running of $Z$ at large cutoff scales, which makes the calculation
increasingly difficult from the numerical point of view with
increasing cutoff scales. However, the most important conclusion from
this section remains the approximate agreement of pole and curvature
masses at vanishing temperature in the fully iterated result.
\begin{table}[t]
	\centering
	\begin{tabular}{cccc}
          \toprule
          \symhspace{1mm}{step} & \symhspace{1mm}{$m_\text{cur}$ [MeV]} & \symhspace{1mm}{$m_\text{pol}$ [MeV]} & \symhspace{1mm}{$\sigma_\text{min}$ [MeV]}\\
          \midrule
	  0 & 412.8 & 412.8 & 16.8 \\
	  1 & 144.8 & \num{142+-2} & 83.5 \\
	  2 & 136.4 & \num{135+-2} & 91.8 \\
	  3 & 135.1 & \num{134+-2} & 93.1 \\
	  4 & 134.9 & \num{133+-2} & 93.2 \\
	  5 & 134.9 & \num{133+-2} & 93.2 \\
          \bottomrule
	\end{tabular}
	\caption{Similar to Tab.~\ref{tab:results:massesT0_tune_polLambda500} but for $\Lambda_\text{UV}=\SI{700}{MeV}$. LPA$'$ masses: \SI{135.2}{MeV}.}
	\label{tab:results:massesT0_tune_polLambda700}
\end{table}

\begin{table}[t]
	\centering
	\begin{tabular}{cccc}
          \toprule
          \symhspace{1mm}{step} & \symhspace{1mm}{$m_\text{cur}$ [MeV]} & \symhspace{1mm}{$m_\text{pol}$ [MeV]} & \symhspace{1mm}{$\sigma_\text{min}$ [MeV]}\\
          \midrule
          0 & 817.0 & 817.0 &5.1\\
          1 & 163.4 & \num{158+-2}&67.9 \\
          2 & 138.5 & \num{137+-2}&89.9\\
          3 & 136.5 & \num{135+-2}&92.4 \\
          4 & 135.4 & \num{134+-2}&93.6\\
          5 & 135.3 & \num{134+-2}&93.6 \\
          \bottomrule
	\end{tabular}
	\caption{Similar to Tab.~\ref{tab:results:massesT0_tune_polLambda500} but for $\Lambda_\text{UV}=\SI{900}{MeV}$. LPA$'$ masses: \SI{134.1}{MeV}.}
	\label{tab:results:massesT0_tune_polLambda900}
\end{table}

\begin{figure}[t]
	\centering
	\includegraphics[width=0.97\columnwidth]{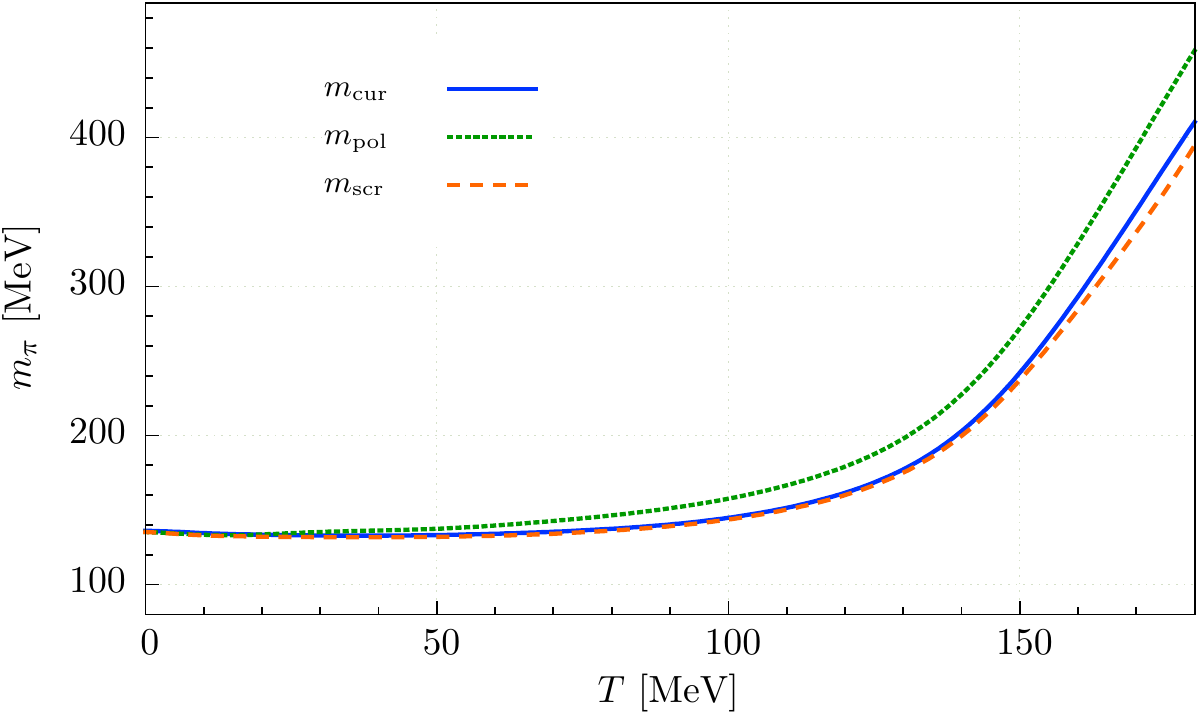}
	\caption{Temperature dependence of different pion mass
          definitions extracted from fully iterated result at $\Lambda_\text{UV}=\SI{1.4}{GeV}$.}
	\label{fig:results:massesQMT_converged}
\end{figure}

At finite temperature, pole and screening masses start to deviate as
expected from \eq{eq:ratiopolescr}, see
Fig.~\ref{fig:results:massesQMT_converged} for the pion masses. For a
similar observation in the NJL model see e.g.\
\cite{Florkowski:1993br}. All low-energy effective models have in
common that they have a {\it physical} UV scale above which they loose
predictive power. In the present formulation this scale is given by
the initial cutoff scale $\Lambda_{\rm UV}$. The thermal range
$\Lambda_{T}$ of the model is defined as the minimal cutoff scale
$\Lambda_{\rm UV}$ above which thermal fluctuations do not probe the
cutoff scale. This is investigated in App.~\ref{app:thermalrange}
and leads to $\Lambda_{T}\lesssim 7\, T$ for the regulators used, see
\eq{eq:R} with $m=2$. 

Here we present results for temperatures $T\lesssim \SI{180}{MeV}$ which is
well covered by $\Lambda_{\rm UV}=\SI{1.4}{GeV}$. The fact that the
curvature mass stays close to the screening mass at all temperatures
is related to the fact that we chose $\bar Z=Z_\perp$ to renormalise
fields and is once again an expression for small non-trivial momentum
dependencies.

The approximate agreement of pole and curvature masses demonstrated
above is a consequence of the momentum dependence obtained from the
converged iteration procedure. In particular, their difference allows
to pin down effects of scale mismatches in existing calculations in
simple but commonly used truncation schemes such as the LPA.  Here we
aim at quantifying the systematic error which is inherent in these
calculations. For different cutoff scales we follow the usual
procedure in the literature and tune initial conditions such that we
obtain correct physical observables in the IR. Then we compute the LPA
onset mass by calculating the momentum-dependent meson propagator
using the given LPA solution, corresponding to the first half of the
first iteration step in Fig.~\ref{fig:iterationScheme}. The pole mass
extracted from this propagator equals, up to approximation effects,
the LPA onset mass, which can be probed directly by including a
coupling to (isospin) chemical potential, see
\cite{Strodthoff:2011tz,Kamikado:2012bt}. This LPA onset mass can now
be compared to the LPA curvature mass extracted from the curvature of
the effective potential as shown in
Tab.~\ref{tab:results:LPAcurvaturepole}. Whereas their ratio tends to
one for smaller UV scales, it increases with the UV scale and reaches
a value of 1.71 for $\Lambda_\text{UV}= \SI{1.4}{GeV}$. 

Such large deviations are in line with earlier studies
\cite{Strodthoff:2011tz,Kamikado:2012bt} where deviations of
\SI{30}{\%} were observed for a 3d regulator function and a cutoff
scale $\Lambda_\text{UV}=\SI{900}{MeV}$. In the present work we use
4d exponential regulators, \eq{eq:R} with $m=2$. For different
regulators the {\it physical} cutoff scales, $k_{\rm phys}(k)$ do not
necessarily agree, for a detailed discussion and applications see
\cite{Pawlowski:2005xe,Marhauser:2008fz}. This entails that coinciding \label{cutoff_ratios}
physical cutoff scales are obtained for different regulator scales
$k$. A rough estimate for this ratio of cutoff scales is given by the
ratio of the (bosonic) flow of the masses. For the standard exponential
regulator \eq{eq:R} with $m=1$ we find $ k_{3\dd}/k_{4\dd,m=1}\approx
3/2$, for the exponential regulator \eq{eq:R} with $m=2$ we find $
k_{3\dd}/k_{4\dd,m=2}\approx 5/4$.

In summary this entails that the commonly used UV cutoff scales
$\Lambda_{\text{UV},3\dd}=\SI{700}{MeV}$ and \SI{900}{MeV} correspond to UV cutoff scales
$\Lambda_{\text{UV},4\dd,m=2}=\SI{560}{MeV}$ and \SI{720}{MeV}, respectively.
For the following numerical examples we will therefore focus on the case of $\Lambda_{\text{UV},4\dd,m=2}=\SI{700}{MeV}$.

\begin{table}[t]
	\centering
	\begin{tabular}{cccc}
          \toprule
          \symhspace{1mm}{$\Lambda_\text{UV}$ [GeV]} & \symhspace{1mm}{$m_\text{cur}$ [MeV]} & \symhspace{1mm}{$m_\text{ons}$ [MeV]} & \symhspace{1mm}{$m_\text{cur}/m_\text{ons}$} \\
          \midrule
          0.5 & 135.0 & \num{109+-2}& 1.24\\
          0.7 & 135.2 & \num{98+-2}& 1.38\\
          0.9 & 135.0 & \num{90+-2}& 1.50\\
	  1.1 & 135.4 & \num{85+-2} & 1.59\\
	  1.4 & 135.3 & \num{79+-2} & 1.71\\
          \bottomrule
	\end{tabular}
	\caption{Comparison of LPA curvature and onset masses for fixed curvature masses in the IR and different UV cutoff scales $\Lambda_\text{UV}$.\vspace{-0.5em}}
	\label{tab:results:LPAcurvaturepole}
\end{table}

At first sight, deviations between the curvature and the onset mass might seem irrelevant for studies at
a given expansion order such as the commonly employed zeroth order
derivative expansion or LPA. However, as already explained in the
beginning of this section, the curvature mass and the ratio
$Z_\perp/Z_\parallel$ sets the relevant scales for quantum and thermal
fluctuations while the pole/onset mass is that of density
fluctuations. In other words, an approximation scheme where these mass
scales differ by \SI{38}{\%} leads to a significant quantitative
change of the ratio of critical temperature $T_c$ over onset chemical
potential $\mu_c$. A rough estimate, assuming that the onset chemical potential/ critical temperature
measured in the respective mass scales stays constant in the different approximation schemes, provides 
\begin{align}\label{eq:ratioTmu}
\left[\0{ \mu_c}{T_c}\right]_{\rm full} / \left[\0{ \mu_c}{T_c}
\right]_{\rm LPA}\approx \left[\0{m_{\rm cur}}{m_{\rm ons}}\right]_{\rm LPA}\approx  1.38\,, 
\end{align}
where the subscript ${}_{\rm full}$ refers to the present
momentum-dependent approximation.

This mismatch of scales has consequences for the scale setting procedure
in LPA. The commonly used procedure is to
fix the physical parameters at the initial UV scale $\Lambda_\text{UV}=\SI{700}{MeV}$ such,
that the pion (pole) mass of approximately \SI{135}{MeV} is the
curvature mass $m_{\pi,\rm cur}$, as it agrees with the pole mass in
this order of the derivative expansion (classical dispersion). And
indeed the full results of the last section justify this
identification. As the scale of quantum and thermal fluctuations is
identical in this approximation due $Z_\perp/Z_\parallel=1$, and
$m_{\rm cur}/m_{\rm scr}=1$ all these fluctuations are treated
self-consistently. In turn, the scale for density fluctuations in LPA is set by the onset mass, see
Tab.~\ref{tab:results:LPAcurvaturepole}, which is approximately \SI{98}{MeV} at a UV cutoff scale $\Lambda_\text{UV}=\SI{700}{MeV}$. 
This entails that the strength of density fluctuations is overestimated by \SI{28}{\%}. 
Alternatively, one can identify the pion mass with the onset mass \cite{Strodthoff:2011tz,Kamikado:2012bt}. With hindsight this comes at the expense of
having a too large scale for quantum and thermal fluctuations of about
\SI{186}{MeV} instead of \SI{135}{MeV}, see
Tab.~\ref{tab:results:LPAcurvaturepole}. In other words, quantum and
thermal fluctuations are underestimated by \SI{38}{\%}. 

To summarise, there is no way of circumventing a mismatch of fluctuation scales
in a truncation scheme with vastly different curvature and onset masses such as the LPA. 
Its implications for the chiral phase boundary at finite density are discussed in
Sec.~\ref{sec:FiniteDensity}. Apart from the quantitative change of
the phase boundary this mismatch may also inflict qualitative changes
at large chemical potential $\mu>\mu_c$, for example if the phase
structure at nuclear densities and beyond involves competing order
effects.

\subsection{Momentum dependence and initial conditions}

In order to study the impact of fully momentum-dependent propagators
on FRG calculations in low-energy QCD, we compare the temperature
dependence of the quark condensate $\chevron{\sigma}$ in different
orders of the truncation scheme.  $\chevron{\sigma}(T)$ is sensitive to 
a correct relative inclusion of thermal and quantum fluctuations as well as 
the absolute scale. We compare three different
truncations, namely LPA, LPA$'$ and the fully momentum-dependent
calculation. 

As already discussed in Sec.~\ref{sec:effectiveaction}, there are two
possibilities to fix the initial conditions. Firstly, one can derive
the initial conditions from computing QCD-flows for the model's
parameters in the UV. In the following we evaluate the consequences of this
set-up by comparing different simpler truncation schemes to the momentum-dependent 
calculation put forward in this work, which is expected to lie closest to the full QCD flow.
 Secondly, one can tune the model parameters such
that the vacuum physics of QCD is reproduced within the respective model and approximation
scheme. In low-energy
QCD without inherent approximations these two sets of initial
conditions agree. In turn, within approximations, they are different.
Hence, we shall consider both approaches in our investigation separately.

\subsubsection{Fixed microphysics} \label{subsec:results:sameUV}

\begin{figure}[t]
	\centering
	\includegraphics[width=0.97\columnwidth]{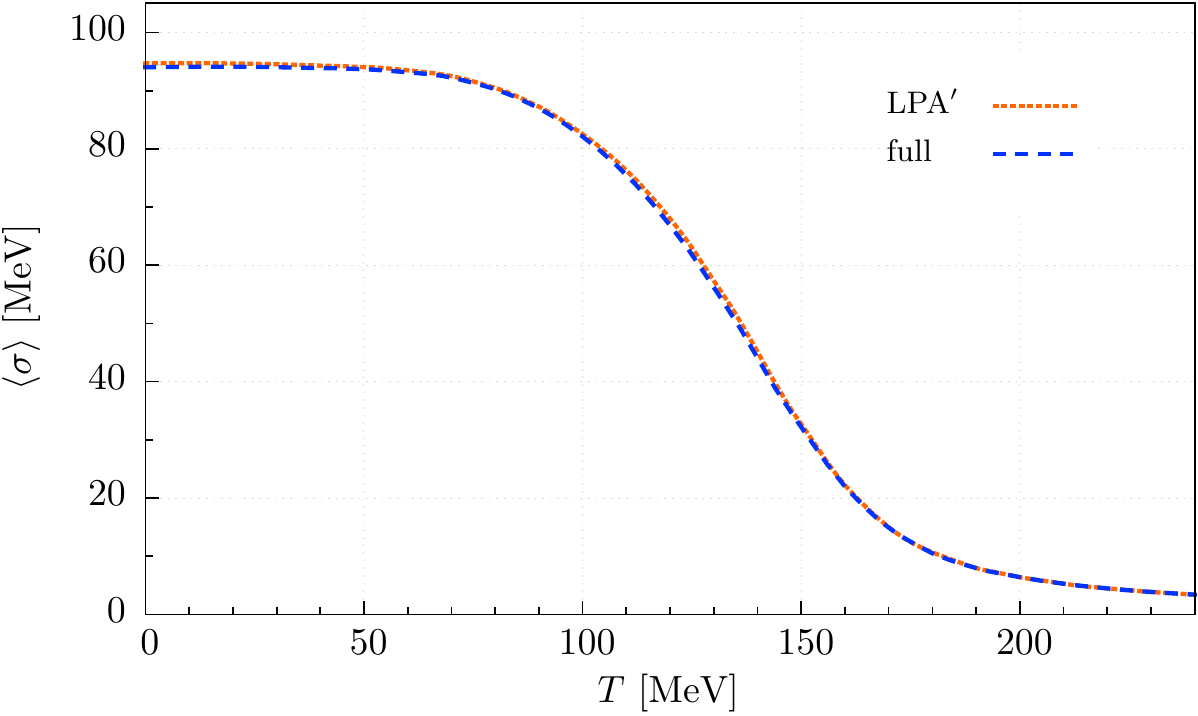}
	\caption[Crossover with fixed UV parameters]{Chiral condensate
          vs.~temperature for fixed UV parameters using different
          truncations and $\Lambda_\text{UV}=\SI{900}{MeV}$.}
	\label{fig:results:sameUVcrossover}
\end{figure}

We first study the effects of including fully momentum-dependent
propagators while keeping the UV-physics of the model fixed at
$\Lambda_\text{UV}=\SI{900}{MeV}$. In such a QCD-embedded approach the input
parameters at $\Lambda_\text{UV}$ are derived within QCD flows. However, in this work, we employ
a parameter set which leads to correct physical observables in the IR
for the full calculation to mimic the effect of fixing initial conditions
from full QCD. Our findings for the chiral crossover are summarised in
Figs.~\ref{fig:results:sameUVcrossover}.

On the one hand, note that for a cutoff scale of $\Lambda_\text{UV}=\SI{900}{MeV}$ no comparison
to the LPA is possible because the LPA calculation with initial
conditions from the full calculation only shows chiral symmetry
breaking for $\Lambda_\text{UV}<\Lambda^*_{\text{LPA}}\approx\SI{600}{MeV}$, see
App.~\ref{app:uvcutofflpa}. Therefore, we restrict ourselves in the LPA
case to a comparison at vanishing temperature for a UV cutoff scale
$\Lambda_\text{UV}=\SI{500}{MeV}$, which can be inferred from
Tab.\ref{tab:results:massesT0_tune_polLambda500}. Note that with the
thermal range $\Lambda_T\lesssim 7 \,T$ we only have access to temperatures
$T\lesssim \SI{70}{MeV}$ anyway, see App.~\ref{app:thermalrange}. 
However, even for $\Lambda_\text{UV}=\SI{500}{MeV}$ LPA is not quantitatively consistent with
the full calculation with deviations of \SI{50}{\%} in $\chevron{\sigma}$ and
$m_\pi$. 

On the other hand, Fig.~\ref{fig:results:sameUVcrossover} shows that the LPA$'$ 
calculation is even quantitatively consistent with the full result. The largest relative
deviations of about \SI{3}{\%} arise in the vicinity of the
pseudo-critical temperature $T_c$ and are certainly related to
pseudo-critical fluctuations.  At about $T_c$, the correlation length
is large, and the system changes its dynamical degrees of freedom from
quarks to mesons. Both properties imply that this region is most
sensitive to momentum fluctuations. This is also in line with the
expectation that including the full momentum dependence or higher
orders of the derivative expansion is crucial around the critical
temperature $T_c$, \eg in order to calculate critical exponents at
high numerical precision, see e.g.\
\cite{Litim:2010tt,Benitez:2011xx}.

  The generally very good agreement of LPA$'$ and the fully momentum-dependent
  truncation in Fig.~\ref{fig:results:sameUVcrossover} is intimately related to
  the use of a cutoff function $R_k$, which regularises both
  frequencies $p_0$ and spatial momenta $\vec{p}$, see
  \cite{Fister:2011uw,Boettcher:2013kia}. Thus, the associated RG
  flows are local both in $\absVal{\vec{p}}$- and in $p_0$-space and
  the argument from the zero temperature discussion in App.~\ref{app:lpaprime}
  applies. Conversely, if we employed a regulator which only affects
  spatial momenta, such as is commonly done in finite-temperature FRG
  calculations, the entire Matsubara summation would be required to
  compute a given RG flow. Put differently, the flows at every scale
  $k$ would receive contributions from both very small and very large
  Matsubara frequencies, thereby at least partly invalidating the
  $T=0$ reasoning presented in App.~\ref{app:lpaprime}. Accordingly, calculations based on
  three-dimensional cutoff functions are anticipated to give rise to
  deviations which are larger than the ones we observe in
  Fig.~\ref{fig:results:sameUVcrossover}.

\subsubsection{Fixed vacuum physics} \label{subsec:results:sameIR}

\begin{table}[b]
	\centering
	\begin{tabular}{ccc||cc}
          \toprule
          step & $m_\text{cur}$ [MeV] & $m_\text{pol}$ [MeV] & $m_\text{cur}$ [MeV]&$m_\text{pol}$ [MeV]\\
          \midrule
          0 & 135.2 & 135.2 & 135.2 & 135.2\\
          1 & 96.5 & \num{96+-2} & 135.5& \num{134+-2}\\
	  2 & 96.8 & \num{96+-2} & 135.5& \num{134+-2}\\
	  3 & 96.8 & \num{96+-2} & 135.6& \num{134+-2}\\
	  4 & 96.8 & \num{96+-2} & 135.6& \num{134+-2}\\
	  5 & 96.8 & \num{96+-2} & 135.6& \num{134+-2}\\
          \bottomrule
	\end{tabular}
	\caption{Pion curvature and pole masses for different 
          iteration steps at $T=0$ applying the iteration procedure on 
          top of a LPA (left) and a LPA$'$ (right) parameter set for a UV cutoff scale $\Lambda_\text{UV} = \SI{700}{MeV}$ . The UV 
          parameters are tuned such that the physical pion mass emerges 
          in the respective zeroth step.}
	\label{tab:results:massesT0_tune_cur}
\end{table}

In the previous section we have discussed the different truncations in
view of the direct connection of the QM model to first-principle
QCD. Then the UV initial conditions can in principle, be calculated
from QCD flows. Within such a combined approach the QM model can be
systematically upgraded to the full low-energy effective
action of QCD. This technically challenging programme is well under
way, and eventually will give quantitative reliability to enhance
low-energy effective model computations.

However, we might also disregard the direct connection to QCD and fix
the initial conditions in the infrared by adjusting the correct vacuum
physics: choose some generic set of (renormalised) IR observables
$(\bar{f}_\pi,\bar{m}_\pi,\bar{m}_\psi)$ and then tune the
microphysics separately in each truncation scheme such that the given
mass scales emerge in the limit $k\to 0$ at vanishing temperature and density. On
the basis of this adjustment one then can study finite-temperature or
finite-density physics. This approach is the standard effective model
approach to low-energy QCD. The discussion in the present section is
meant to evaluate and improve the reliability of this set-up. 

Note first that such a procedure falls short of a direct connection to
first-principle QCD in the UV. Moreover, in particular in low-order
approximations such as LPA, some fluctuation physics is simply stored
in the initial condition. For example, in LPA for the model at hand,
we have to change the UV initial conditions such that they effectively
take care of the missing momentum effects. In order to assess the
impact of adding the full momentum dependence on top of a given LPA or
LPA$'$ solution, we show in Tab.~\ref{tab:results:massesT0_tune_cur}
the iteration procedure applied to a given LPA or LPA$'$ solution,
referred to as zeroth iteration step in
Tab.~\ref{tab:results:massesT0_tune_cur}.  On the one hand, as
expected from Sec.~\ref{sec:results:massesandfluctuations}, the
iteration on top of the given LPA result shows a large deviation of
over \SI{40}{\%} in the masses after the first iteration step,
illustrating again the large mismatch of fluctuation scales in the
LPA. On the other hand, with a deviation of less than one percent, the
LPA$'$ solution is already very close to the full result.  Even more
conveniently for practical purposes, the first iteration step deviates
less than one per mill from the full result. Hence, evaluating the
momentum dependence on the basis of a given LPA$'$ solution, already
provides a simple but very good approximation to the full
momentum-dependent solution.

\begin{figure}[t]
	\centering
	\includegraphics[width=0.97\columnwidth]{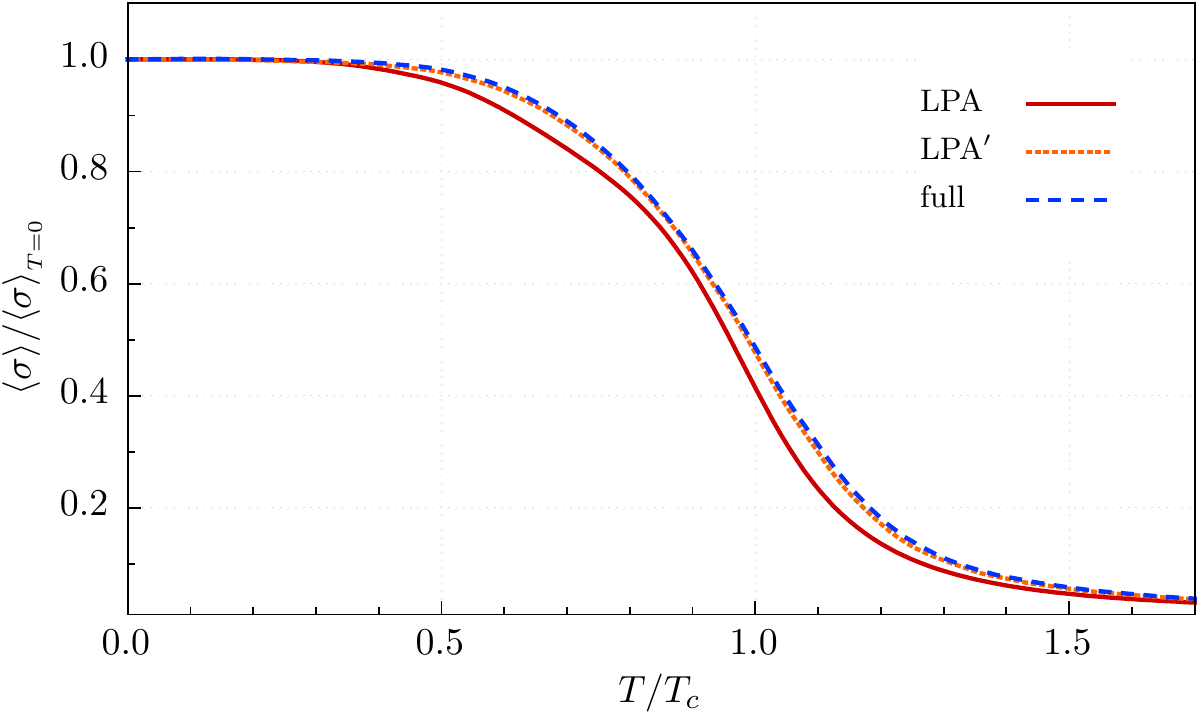}
	\caption[Crossover with fixed IR parameters]{Chiral condensate
          vs. temperature for fixed physical IR parameters in the
          vacuum using different truncations and $\Lambda_\text{UV}=\SI{900}{MeV}$.}
	\label{fig:results:sameIRcrossover}
\end{figure}

Our findings on the chiral crossover are summarised in
Fig.~\ref{fig:results:sameIRcrossover}. We again observe that the LPA$'$
scheme approximates the full flow very well. To be more precise, the relative deviation of the chiral condensate never exceeds $\SI{3}{\%}$. Note, however, that the
relative deviation again exhibits a peak centered at the
pseudo-critical temperature, which indicates accuracy issues in the
presence of pseudo-critical fluctuations. Possible reasons for the generally good agreement between the two schemes were already extensively studied
in the preceding paragraph as well as in App.~\ref{app:lpaprime}.
Accordingly, we will concentrate on the remaining
comparison between the LPA and the full calculation.

Here, Fig.~\ref{fig:results:sameIRcrossover} reveals that despite 
large deviations of about \SI{19}{\%} in $T_c$ measured in absolute scales,
which is to large parts caused by different sigma mass ranges which 
can be reached in the different approximation schemes, the deviations
in terms of relative scales are rather small as well, at least for
temperatures outside the critical regime. This presumably reflects the
fact, that the propagators' non-trivial momentum dependencies enter
the computation of the chiral condensate from the effective potential
only indirectly. Put differently, the crossover's shape seems to be
mainly fixed by the infrared mass scales in the vacuum. 

\begin{figure}[t]
	\centering
	\includegraphics[width=0.97\columnwidth]{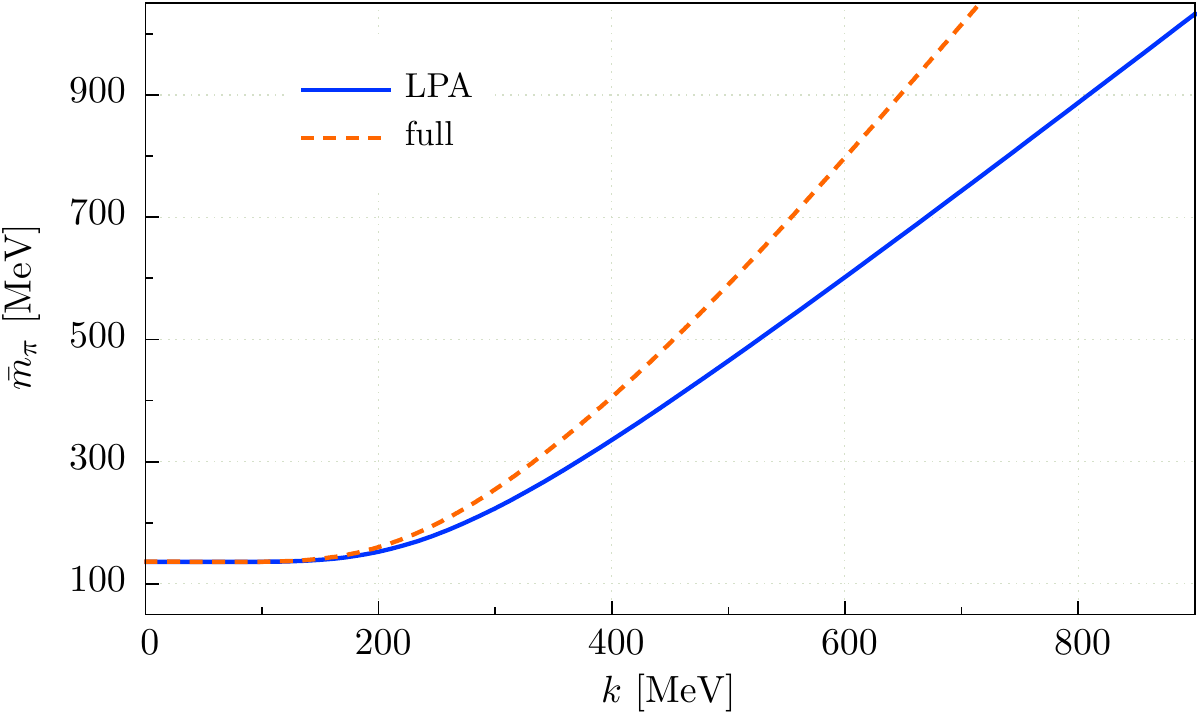}
	\caption{Scale dependence of the pion mass at $T=0$ in LPA and in the full calculation for $\Lambda_\text{UV}=\SI{900}{MeV}$ for fixed physical IR parameters.}
	\label{fig:results:scaledepmasses}
\end{figure}
The quantitative discrepancy between LPA and the full calculation can be understood
most easily from Fig.~\ref{fig:results:scaledepmasses}, where we compare 
the scale dependence of the pion mass for fixed initial
conditions in the IR as an illustration of cutoff regions where the LPA calculation gives qualitatively or quantitatively correct results compared to the
full calculation. Remarkably, both is only the case for cutoff scales below \SI{200}{MeV} whereas for larger scales the results are not even qualitatively
correct as it is clearly visible from the different slopes as a result of the fixed renormalised Yukawa coupling in the full calculation.

\subsection{Finite density}\label{sec:FiniteDensity}  

Let us finally evaluate the consequences of the results in the last
two sections for finite density computations. These observations point
at a mismatch of scales in LPA between quantum/thermal and density
fluctuations with a factor as given in \Eq{eq:ratioTmu}. As before
we employ the cutoff scale $\Lambda_\text{UV}=\SI{700}{MeV}$ as 
numerical example.

To illustrate its consequences we consider a simple application to the
physics of the phase diagram in LPA, where we resort to the simple
rescaling argument that has worked so successfully for quantum and
thermal fluctuations in Section~\ref{subsec:results:sameIR}. Within
this reasoning, the overestimation of density fluctuations can be
approximately undone by simply rescaling the chemical potential axis
with a factor $1.38$. This weakens the curvature of the chiral phase
boundary $T_c(\mu)/T_c(0)$ at finite chemical potential. At small
chemical potential the phase boundary can be expanded in powers of
$\mu^2$ as follows:
\begin{align}\label{eq:curvf}
  \frac{T_c(\mu)}{T_c(0)} = 1-\kappa_\mu \left(\frac{\mu}{\pi T_c(0)}
  \right)^2+\mathcal{O}\biggl(\!\left(\frac{\mu}{\pi
      T_c(0)}\right)^4\biggr)\,,
\end{align}
for a discussion of the phase structure of the $N_f=2$ quark-meson
model in LPA and LPA$'$ as well as with higher order quark-meson scattering
processes see \cite{Pawlowski:2014zaa}. \Eq{eq:curvf} entails that a
stretching of the chemical potential axis with a factor $1.38$ weakens
the curvature $\kappa_\mu$ by a factor $0.53$. In
\cite{Pawlowski:2014zaa} the curvature was computed from the chiral
susceptibility as $\kappa_{\mu,\rm LPA}\approx 1.4$. Applying the reduction factor, this reduces to $\kappa_\mu\approx 0.74$, 
which influences the result in the direction of the lattice curvature $\kappa_\mu\approx 0.5$ \cite{deForcrand:2002ci}. 

The above discussion suggests that taking into account the full
momentum dependence of the propagators in the quark-meson model may
account for the mismatch between the curvature of the phase boundary
computed in the models and the lattice result. However,
\cite{Pawlowski:2014zaa} also contains a computation with constant
wavefunction renormalisations for meson and quark fields (LPA$'$) and a fully
(meson-)field-dependent Yukawa coupling (corresponding to higher order 
quark-mesonic scattering processes). The
curvature $\kappa_\mu $ for this computation agrees surprisingly well
with the LPA result. This means that either the higher order
quark-meson scatterings counterbalance the momentum effects on the
curvature, or the LPA$'$ computation with a 3d regulator in
\cite{Pawlowski:2014zaa} does not cover the full momentum dependence. 
The results for this investigation will be presented elsewhere.

\section{Conclusion}

In the present work we have discussed the relation between different
mesonic mass scales in low-energy QCD within a $N_f=2$ quark-meson
model. To that end we have computed fully momentum-dependent two-point
functions of pions and the $\sigma$-meson as well as a full mesonic
potential within a functional renormalisation group approach. This
allows us to compare pole, screening and curvature masses both at
vanishing and finite temperature; respective definitions are discussed
in detail in Section~\ref{sec:massdefinitions}. Whereas pole and screening mass coincide by
definition for vanishing temperature they start to deviate at finite
temperatures. Moreover, we find that the fluctuation scales for
density fluctuations, related to the pole masses at vanishing
temperature, and that for quantum and thermal fluctuations, related to
the curvature masses, almost agree.

The present momentum-dependent set-up has also been used to evaluate
the reliability of lower order approximations. In the present work we
considered the two standard approximations to the QM model: the local
potential approximation (LPA), where only classical propagators and
the full mesonic potential are considered, and LPA$'$, which
additionally involves constant wavefunction renormalisations for the
mesonic fields. Our results show a very good agreement between the
fully momentum-dependent calculation and the LPA$'$ calculation with
relative deviations of at most \SI{3}{\%} in a narrow region around
the pseudo-critical temperature, which justifies the use of the LPA$'$
as simple but very reliable truncation which includes already a large
part of the momentum dependence of the full propagator.

In turn, we observe a large mismatch between the
pion onset and curvature masses in LPA at vanishing temperature which
reaches \SI{38}{\%} for typical UV cutoff scales $\Lambda_\text{UV}=\SI{700}{MeV}$,
which has important implications for the relative fluctuation scales
for vacuum/thermal and density fluctuations. Neglecting
these effects leads to large systematic errors at finite chemical
potential. Moreover, even at very low UV cutoff scales the LPA truncation for fixed
initial conditions in the UV does not lead to quantitatively correct 
results compared to the outcome of the full calculation.

In the line of these findings we estimated the effect of this mismatch
of quantum/thermal versus density fluctuation scales in LPA within a
simple rescaling the chemical potential axis accordingly, see
Section~\ref{sec:FiniteDensity}. This simple argument leads to a
result for the curvature of the chiral phase boundary, which lies
reasonably close to the lattice result, but remains to be checked in
larger truncation schemes.

An investigation of the combination of the approximation in
\cite{Pawlowski:2014zaa} with $O(4)$-symmetric regulators and full
momentum dependence at finite density will be presented elsewhere.
This requires the extension of the present 4d regulator classes to
finite chemical potential, which is also tightly linked to the
computation of real time quantities such as spectral functions in a
fully $O(4)$ and Minkowski-invariant set-up, which will be discussed in
a future publication. The present findings
strongly emphasise the necessity of such a symmetry-preserving
approach.

\acknowledgments We thank L.~Fister, F.~Rennecke and L.~von Smekal for
discussions. JMP thanks the Yukawa Institute for Theoretical Physics,
Kyoto University. Discussions during the YITP workshop YITP-T-13-05 on
'New Frontiers in QCD' were useful to complete this work. This work is
supported by the Helmholtz Alliance HA216/EMMI and the grant
ERC-AdG-290623.

\appendix

\section{Truncation and flow equations} \label{app:flow_equations} 

In this Appendix we discuss details on the truncation scheme as well as
explicit expressions for the flow equations for the effective
potential and the two-point function.

The flow equation for the momentum-dependent part of the two-point
function involves the full three-point vertices, which in turn carry a
non-trivial momentum dependence.  At this point we completely neglect
the momentum dependence of these vertices and simply determine them
from the effective potential. Note in particular that the iteration
procedure only involves the difference $\Delta\Gamma^{(2)}_{k}(p)$
between the two-point function evaluated at finite and at vanishing
external momentum. Note furthermore that the terms proportional to
$Y_{k}$ even in our expansion lead to momentum-dependent three- and
four-point vertices.  Taking into account these terms in the
calculation will lead to a $\Lambda^{2}$-rise instead of the correct
$\Lambda^{-2}$-decay from the factorization property which is linked
to the fact that the derivative expansion only works well for
$p^{2}/k^{2} \ll 1$. We want a procedure which is correct at vanishing
and at asymptotically large momenta where in both cases the $Y_{k}$
contribution to the three-point vertices is absent and is expected to
be subleading also for intermediate momenta. Therefore, we determine
the mesonic three-point vertices solely from the effective potential,
for explicit expressions see \cite{Kamikado:2013sia,Tripolt:2013jra}.
The full inverse two-point function $\Gamma^{(2)}$ is obtained via
\begin{equation}
\Gamma^{(2)}(p)=\Delta \Gamma^{(2)}(p)+ U^{(2)},
\end{equation}
where $U^{(2)}$ denotes the appropriate second field derivative of the
effective potential. As pointed out in Sec.~\ref{sec:massdefinitions}, the
momentum-dependent propagator now allows to calculate different particle masses. 
Thereby the calculation of screening, pole and onset masses requires
analytical continuation, which is here performed by means of Chebyshev
approximation. For the purpose of estimating the corresponding error, we have varied 
the Chebyshev approximation's order $N_\text{cheb}$ over a
range of $N_\text{cheb} \in \{50, \ldots , 100\}$.

Next we specify the explicit expressions for the flow equations for the
effective potential and the two-point function. Here, we employ
4d regulator functions of the form
\begin{subequations}
\begin{align}\label{eq:Rk}
R_k^{B/F}(q^2)  = \Delta \Gamma^{(2)}_{B/F,k}(q^2,\exppt) 
\cdot r(q^2/k^2) \fineq{.}
\end{align} 
All practical calculations are performed using an exponential
regulator shape function, 
\begin{align}\label{eq:rm}
r(x)=x^{m-1}/(e^{x^m}-1)\,,
\end{align} 
\label{eq:R}\end{subequations}
with $m=2$. The
equation for the effective potential reads
\begin{equation}
	\begin{split}
          \partial_t U_k ={}& \tfrac{1}{2} I_{B,k}^{(1)}(m_\sigma^2)+
          \tfrac{N-1}{2}I_{B,k}^{(1)}(m_\pi^2) \\
          & - 4 N_c N_f I^{(1)}_{F,k}(m_\psi^2) ,
	\end{split}
	\label{eq:U_flow}
\end{equation}
where $m_\pi^2=2U'$, $m_\sigma^2=2U'+4 U''\rho$ and $m_\psi^2=h^2
\rho$. The flow equations for the inverse pion and sigma meson
two-point functions are given by \cite{Kamikado:2013sia}
\begin{widetext}
	\begin{align}
		\begin{split}
			\del[t] \Delta \GammaTwo_{\pi, k}(p^2, \rho) ={}& \rho 
\, \bigl( 4 U'' \bigr)^2 \cdot \Bigl( \Delta J_{B,k}(p^2;m_\sigma^2,m_\pi^2) 
+ \Delta J_{B,k}(p^2;m_\pi^2,m_\sigma^2) \Bigr) \\
			& -8 N_c N_f h^2 \cdot \Bigl(  \Delta J^{(1)}_{F,k} (p^2) - 
2 m_\psi^2 \Delta J^{(2)}_{F,k}(p^2) + m_\psi^2 \Delta J^{(3)}_{F,k}(p^2) 
\Bigr) \fineq{,}
		\end{split}\\
		\label{eq:floweq_sigma}
		\begin{split}
                  \del[t] \Delta \GammaTwo_{\sigma, k}(p^2, \rho) ={}&
                  \rho \, \bigl( 12 U'' + 8 \rho U''' \bigr)^2 \cdot
                  \Delta J_{B,k}(p^2;m_\sigma^2,m_\sigma^2) + (N-1) \,
                  \rho
                  \, \bigl( 4 U'' \bigr)^2 \cdot \Delta J_{B,k}(p^2;m_\pi^2,m_\pi^2) \\
                  & -8 N_c N_f h^2 \cdot \Bigl( \Delta J^{(1)}_{F,k}
                  (p^2) + 2 m_\psi^2 \Delta J^{(2)}_{F,k}(p^2) +
                  m_\psi^2 \Delta J^{(3)}_{F,k}(p^2) \Bigr) \fineq{.}
		\end{split}
	\end{align}
\end{widetext}
Here, we have defined $\Delta J_k(p^2) = J_k(p^2) - J_k(0)$. Note that
the first lines of the flow equations
\eqref{eq:U_flow}\textendash\eqref{eq:floweq_sigma} can be identified
with bosonic contributions, whereas the second lines constitute
fermionic ones. Finally, the functions occurring on the right-hand sides of the above
flow equations are given by
\begin{align*}
		I_{\pi,k}^{(1)}(m^2) &= \sumint_q \frac{\del[t](\Delta
 \GammaTwo_{\pi, q} r_q)}{\Delta 
\GammaTwo_{\pi, q} (1 + r_q) + m^2} \fineq{,} \\
		I_{\sigma,k}^{(1)}(m^2, \rho) &= \sumint_q 
\frac{
\del[t](\Delta \GammaTwo_{\sigma, q} r_q)}{\Delta 
\GammaTwo_{\sigma, q} (1 + r_q) + m^2 + (\rho - 
\exppt) Y_q q^2} \fineq{,} \\
		I_{F,k}^{(1)}(m^2) & = \sumint_q \frac{q^2 
\dot{r}_q(1+r_q)}{q^2(1+r_q)^2+m^2} \fineq{.}
\end{align*}
We employ the abbreviation $r_q \equiv r(q^2/k^2)$. Analogously, we
have written $Y_q$ instead of $Y_k(q^2)$ and similar for $\Delta
\smash{\GammaTwo_q}$. In addition, one finds
\begin{align*}
  J_{B,k}(p^2;\! m_A^2,\!m_B^2) &\!=\! \sumint_q\!\! \tfrac{\del[t](\Delta 
\GammaTwo_{\pi, q} r_q)}{ 
\left[ \!\Delta \GammaTwo_q\! (1 + r_q)\! +\! m^2\! \right]^2_A \left[\! \Delta 
\GammaTwo_{p+q}\! (1 + r_{q+p})\! +\! m^2 \!\right]_B} \fineq{,} \\
  J_{F,k}^{(1)}(p^2;m^2)\!&=\!-\!\sumint_q\tfrac{\dot{r}_q(1+r_q)^2(1+r_{p+q}) q^2 
q\cdot(q+p)}{(q^2\!(1+r_q)^2\!+\!m^2)^2((q+p)^2\!(1+r_{q+p})^2\!+\!m^2)} \fineq{,} \\
  J_{F,k}^{(2)}(p^2;m^2)\!&=\sumint_q\tfrac{\dot{r}_q(1+r_q) q^2}{(q^2\!(1+r_q)^2
\!+\!m^2)^2((q+p)^2\!(1+r_{q+p})^2\!+\!m^2)} \fineq{,} \\
  J_{F,k}^{(3)}(p^2;m^2)\!&=\sumint_q\tfrac{\dot{r}_q(1+r_{p+q})
    q\cdot(q+p)}{(q^2\!(1+r_q)^2\!+\!m^2)^2((q+p)^2\!(1+r_{q+p})^2\!+\!m^2)}
  \fineq{.}
\end{align*}
The particle indices $A,B \in \{\pi,\sigma\}$ in the denominator
affect both the inverse propagators and the mass terms.

\section{Iteration procedure}
\label{app:iteration}

The starting point is the calculation of the full effective potential
using classical propagators which is then used as input for the
calculation of the momentum-dependent part of two-point
functions. Since the equation for the two-point function requires
derivatives of the effective potential at field value $\exppt$ as
input, the flow equation for the effective potential is conveniently
solved using a Taylor expansion at a fixed expansion point, see
\cite{Pawlowski:2014zaa} for details. The calculation of the
momentum-dependent propagators and their feedback then constitutes the
first iteration step which is subsequently repeated until convergence
is reached.

Most importantly, we want the expansion point in the final iteration
step to coincide with the (IR-)minimum of the effective potential, as
we restrict ourselves to the zeroth order in the field expansion of
the two-point function. In every iteration step the expansion point is
chosen such that it coincides with the minimum of the effective
potential in that particular iteration step. Subsequently, the two-point
function is then computed using the couplings at the same expansion
point. This construction ensures that the expansion point converges
towards the minimum of the effective potential in the converged result
with a propagator evaluated at the same point. Furthermore only
couplings at the minimum of the effective potential enter the
calculation and one never expands inside the shallow region of the
potential.

\section{Thermal range}\label{app:thermalrange}

Usually the initial conditions are kept temperature-independent. This
is only consistent for sufficiently large initial scales $\Lambda_{\rm
  UV}$. For large cutoff scales they provide the characteristic mass
scale of the (regularised) model. In thermal perturbation theory the
thermal range is given by $\exp(-m/(2 T))$, with $m$ being the mass of
the theory. Hence we expect that thermal fluctuations are suppressed
exponentially with $\exp(-k/(\alpha_\text{reg} T))$ with a
regulator-shape-function-dependent factor $\alpha_\text{reg}$, see
\cite{Fister:Diss,Fister:2011uw,Litim:2006ag} for a discussion of
thermal flows. In Sec.~\ref{sec:results:massesandfluctuations} we have
discussed the relation between cutoff scales for different regulators,
see page~\pageref{cutoff_ratios}, left column. Hence, without the shape dependence of
$\alpha_\text{reg}$ the ratio of thermal ranges for different
regulators should agree with that of the UV cutoff scales discussed in
Sec.~\ref{sec:results:massesandfluctuations}. Below we consider as in
Sec.~\ref{sec:results:massesandfluctuations} the 3d optimized regulator and the
4d exponential regulators with $m=1,2$. The comparison of the latter
provides information about the shape dependence of $\alpha_\text{reg}$.

For a given maximal temperature of interest
$T_\text{max}$, we define a minimal UV cutoff scale \smash{$\Lambda^{(n)}_T$}
by the condition that the UV flow of the coupling $\lambda_n$ stays
approximately temperature-independent above that scale, \ie
\begin{equation}
\label{eq:thrange}
\left|\frac{\dot \lambda_n^{T=T_\text{max}}(k)-\dot \lambda_n^{T=0}(k)}{
\dot \lambda_n^{T=0}(k)}\right|<0.05\quad\text{for}\quad k>\Lambda_T^{(n)}\,,
\end{equation}
where the chosen bound of five percent on the right hand side is
clearly strongly model-dependent. However, we emphasise that the
criterion from above is conservative in the sense that the resulting
deviation in the infrared will be significantly smaller than \SI{5}{\%}. Here
we consider the flows of the two relevant parameters $\lambda_1$ and
$\lambda_2$ corresponding to mass and $\phi^4$-coupling. The cleanest
set-up to investigate this question is that of single field mode flows
where the corresponding dimensionless mass parameter $\omega=m^2/k^2$
is set to zero. This allows to compare the thermal range of
different regulator functions.
\begin{figure}[ht]
	\centering
	\includegraphics[width=0.97\columnwidth]{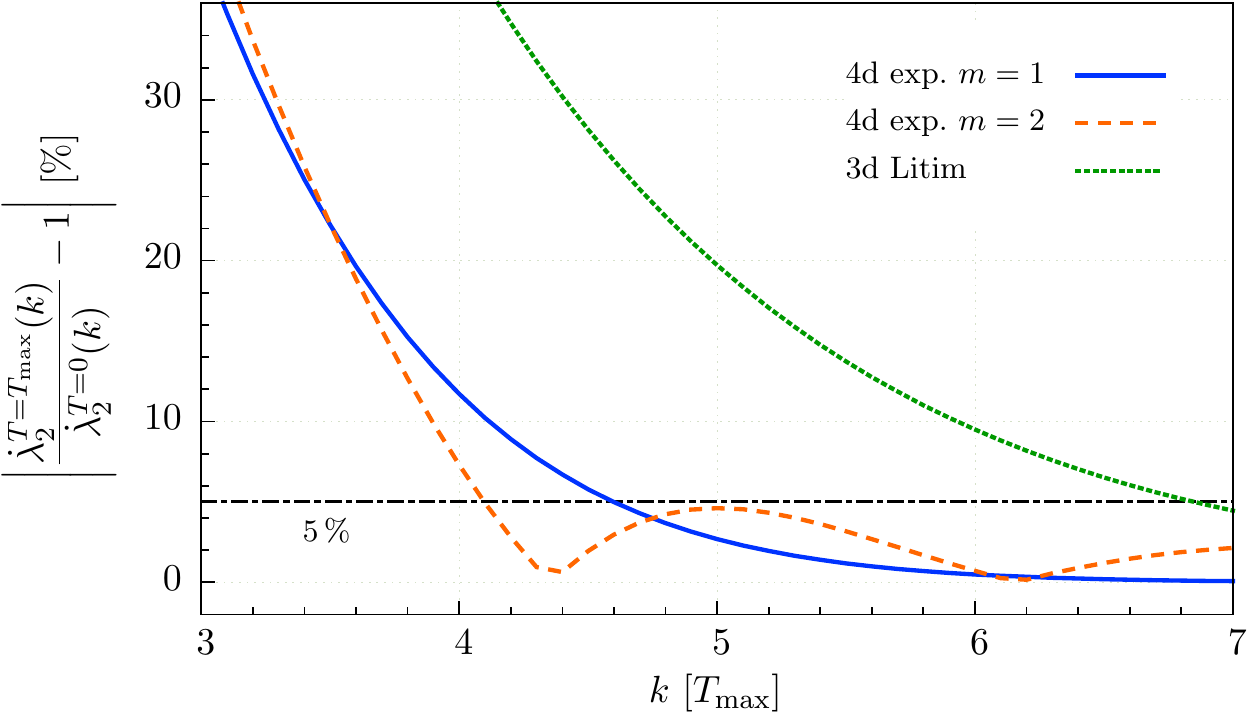}
	\caption{Thermal range for a single bosonic field mode for different regulator functions.}
	\label{fig:app:LambdaT}
\end{figure}

\begin{table}[!ht]
	\centering
	\begin{tabular}{ccccc}
          \toprule
          regulator & \symhspace{0.5em}{$\Lambda^{(1)}_{T,\text{bos}}$} & \symhspace{0.5em}{$\Lambda^{(2)}_{T,\text{bos}}$} & \symhspace{0.5em}{$\Lambda^{(1)}_{T,\text{ferm}}$} & \symhspace{0.5em}{$\Lambda^{(2)}_{T,\text{ferm}}$} \\
          \midrule
          4d exp.~($m=1$)&3.8&4.6&3.1&3.1\\
          4d exp.~($m=2$)&5.5&4.8&6.75&6.75\\
	  3d Litim &5.6& 6.9&5.6&6.9\\
          \bottomrule
	\end{tabular}
	\caption{Thermal range from a single bosonic/fermionic field mode as defined by \eq{eq:thrange} for different regulator functions. All ranges are given in units of the maximal temperature $T_\text{max}$. For $m=2$ an exponential enveloping function was used.}
	\label{tab:app:LambdaT}
\end{table}
Let us recall the (bosonic) cutoff ratios discussed in
Sec.~\ref{sec:results:massesandfluctuations} with $
k_{3\dd}/k_{4\dd,m=1}\approx 3/2$ and $ k_{3\dd}/k_{4\dd,m=1}\approx 5/4$ for
the exponential regulators defined in \eq{eq:R}. For $m=1$ this fits
well to the ratio of thermal ranges
$\Lambda_{T,\text{bos},3\dd}/\Lambda_{T,\text{bos},4\dd,m=1}\approx
3/2$. In turn, for $m=2$ the ratios vary with large uncertainties, and
are structurally smaller than the cutoff ratios. This comes from the
sharp decay of the $m=2$ cutoff function which increases the thermal range,
see \cite{Fister:Diss} for computations for the pressure.

\section{Characteristic scales}\label{app:uvcutofflpa}

\begin{figure}[t]
	\centering
	\includegraphics[width=0.97\columnwidth]{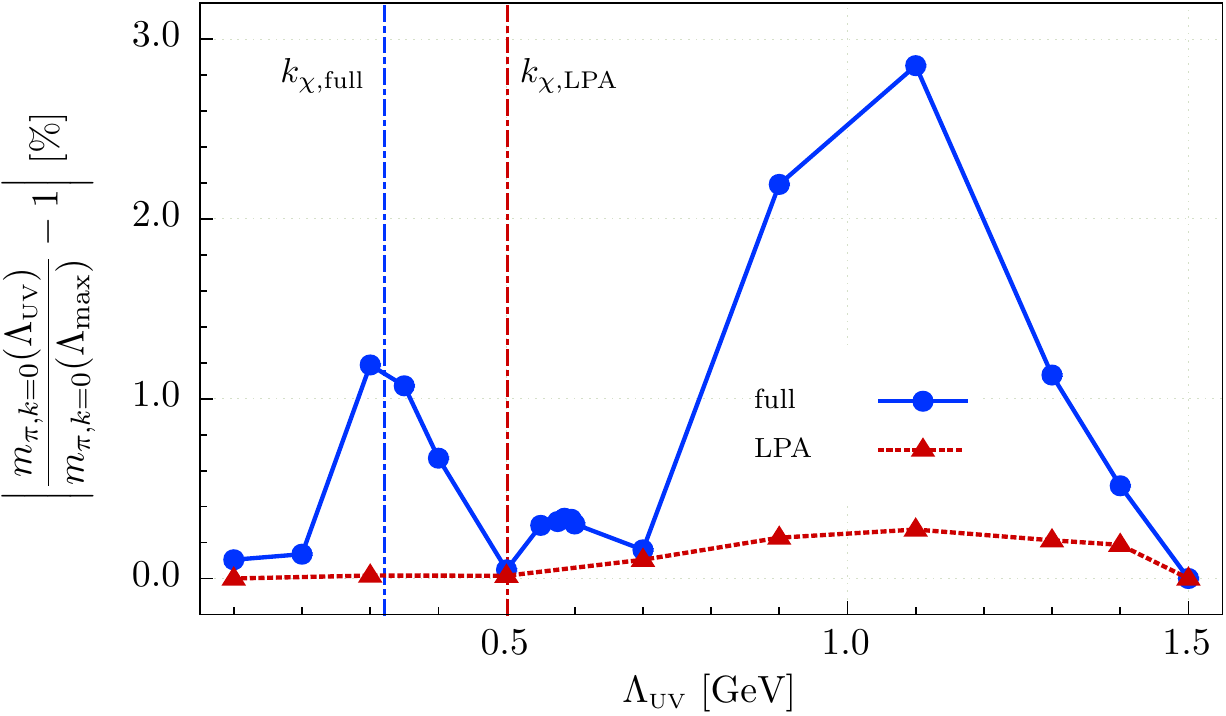}
	\caption{Determination of $\Lambda_\text{cl}$ for LPA and for the full calculation.}
	\label{fig:app:Lambdacl}
\end{figure}

$\Lambda_\text{cl}$ is defined as the
scale until which a description with a classical Lagrangian at the UV
scale in the sense of a $\phi^4$-potential is sufficient in order to
reach the full result in the IR. More concretely, starting from a
given UV scale $\Lambda_\text{UV}$ the full flow is integrated down to
a lower scale $\Lambda_\text{UV}'<\Lambda_\text{UV}$ where a global
rescaling is applied in order to achieve $Z_{\Lambda_\text{UV}'}(0)=1$. The corresponding
rescaled relevant couplings along with a classical propagator are used
as input for a full calculation starting from $\Lambda'_\text{UV}$. We
define $\Lambda_\text{cl}$ as the scale above which the IR observables
such as masses and the minimum of the effective potential do not
deviate significantly from the result of the full calculation
initiated at $\Lambda_\text{UV}$. The calculation for the
determination of $\Lambda_\text{cl}$ is illustrated in
Fig.~\ref{fig:app:Lambdacl}. Interestingly, in LPA the deviation never
exceeds \SI{0.2}{\%}, which is consistent although not equivalent to the
statement that the $\phi^4$ truncation already leads to quantitatively
correct results in LPA. Also in the full calculation the relative
deviation never exceeds \SI{3}{\%} but peaks at \SI{1.1}{GeV} and \SI{0.3}{GeV}, where
the dominant contributions from the fermionic/bosonic flow arise. In
this sense the quark-meson model at vanishing temperature is trivial
as it can be described quantitatively using a classical
potential. This is a consequence of the fermionic contributions to the
flow, whereas it no longer remains true upon increasing the number of
bosons or reducing the number the number of fermions like in the
purely bosonic $O(N)$ model, where also the convergence properties of
the Taylor expansion worsen significantly. The same is true for the
case of finite temperature where higher-order contributions are
required to observe convergence in the Taylor expansion
\cite{Pawlowski:2014zaa}.

\begin{figure}[b]
	\centering
	\includegraphics[width=0.97\columnwidth]{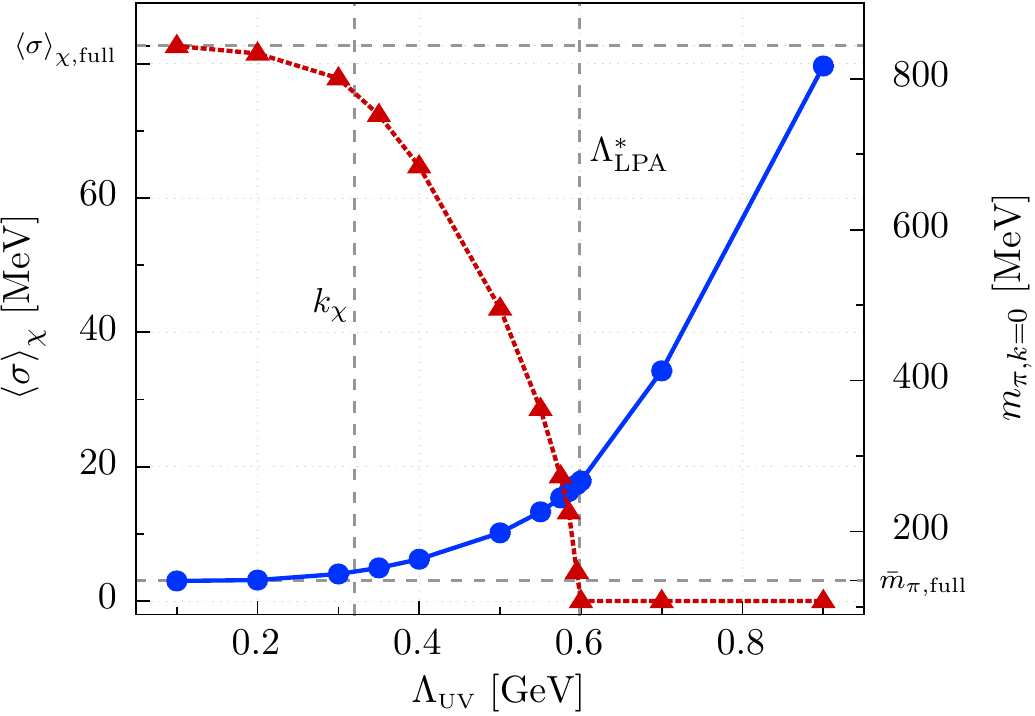}
	\caption{Determination of the scale $\Lambda_\text{LPA}^*$: Pion mass and minimum of the effective potential in the IR as function of the UV cutoff scale $\Lambda_\text{UV}$, where a LPA calculation with initial conditions from the full calculation was initiated.}
	\label{fig:app:LambdaLPA}
\end{figure}
Finally, for a meaningful comparison between LPA and the full calculation, it is insightful to define the scale $\Lambda^*_\text{LPA}$ which is the largest cutoff scale in which an LPA calculation with initial conditions from the full calculation shows chiral symmetry breaking. This puts a natural upper limit on possible UV cutoff scale for LPA calculations using QCD initial conditions. The determination of $\Lambda^*_\text{LPA}$ is illustrated in Fig.~\ref{fig:app:LambdaLPA}, where we plot pion mass and minimum of the
effective potential in the IR as a function of the UV cutoff scale $\Lambda_\text{UV}$, where a LPA calculation with initial conditions from the full calculation was initiated. Only below \SI{600}{MeV}
the LPA calculation shows chiral symmetry breaking, which is the largest scale for which a comparison to the full calculation using fixed UV initial conditions is possible. A more reasonable scale for this comparison which is used in this work is \SI{500}{MeV}, where the chiral condensate in the LPA calculation reaches at least one half of the value from the full calculation. However, the cutoff scale can not be lowered to arbitrarily small scales, where the LPA result will converge towards the result of the full calculation by construction, as this just implies storing the full
flow in the initial condition for the LPA calculation. In any case a lower bound for these fluctuation scales is provided by the the spontaneous symmetry
breaking scale $k_\chi$, which is found at $k_\chi\approx \SI{320}{MeV}$ in the full calculation.

\section{Initial conditions}\label{app:initialcond}

In this appendix, we specify the initial conditions used in the calculations presented in this paper. The UV potential is taken as
\begin{equation}
U_{k=\Lambda_\text{UV}}=a\phi^2+b\phi^4
\end{equation}
and the explicit symmetry breaking term is taken to be $-c\sigma$.

\begin{table}[ht]
	\centering
	\begin{tabular}{ccccc}
		\toprule
        	$\Lambda_\text{UV}$ [GeV] & \symhspace{4mm}{$a/\Lambda_\text{UV}^2$} & \symhspace{5mm}{$b$} & \symhspace{3mm}{$c/\Lambda_\text{UV}^ 3$} & $\sigma_0$ [MeV]\\
		\midrule
		0.5 & 0.608 & 2.446 & 0.0183 & 69.4 \\
		0.7 & 1.021 & 2.334 & 0.0083 & 55.5 \\
		0.9 & 1.304 & 2.055 & 0.0047 & 48.0 \\
		1.4 & 1.795 & 0.505 & 0.0018 & 34.0 \\
		\bottomrule
	\end{tabular}
	\caption{Initial conditions for full calculations with physical IR masses. The renormalised Yukawa coupling is $\bar{h}=3.226$ for all UV cutoff scales.}
	\label{tab:tuningPars_Full}
\end{table}

\begin{table}[!hb]
	\centering
	\begin{tabular}{ccccc}
		\toprule
        	$\Lambda_\text{UV}$ [GeV] & \symhspace{4mm}{$a/\Lambda_\text{UV}^2$} & \symhspace{5mm}{$b$} & \symhspace{3mm}{$c/\Lambda_\text{UV}^ 3$} & $\sigma_0$ [MeV]\\
		\midrule
		0.5 & -0.009 & 5.812 & 0.0136 & 93.0 \\
		0.7 & 0.352 & 4.676 & 0.0049 & 93.0 \\
		0.9 & 0.532 & 3.451 & 0.0024 & 93.0 \\
		1.4 & 0.735 & 0.531 & 0.0006 & 93.0 \\
		\bottomrule
	\end{tabular}
	\caption{Initial conditions for LPA calculations with physical IR masses. The Yukawa coupling is $h=3.226$ for all UV cutoff scales.}
	\label{tab:tuningPars_LPA}
\end{table}

\section{Convergence properties} \label{app:convergence}

\begin{figure}[t]
	\centering
	\includegraphics[width=0.97\columnwidth]{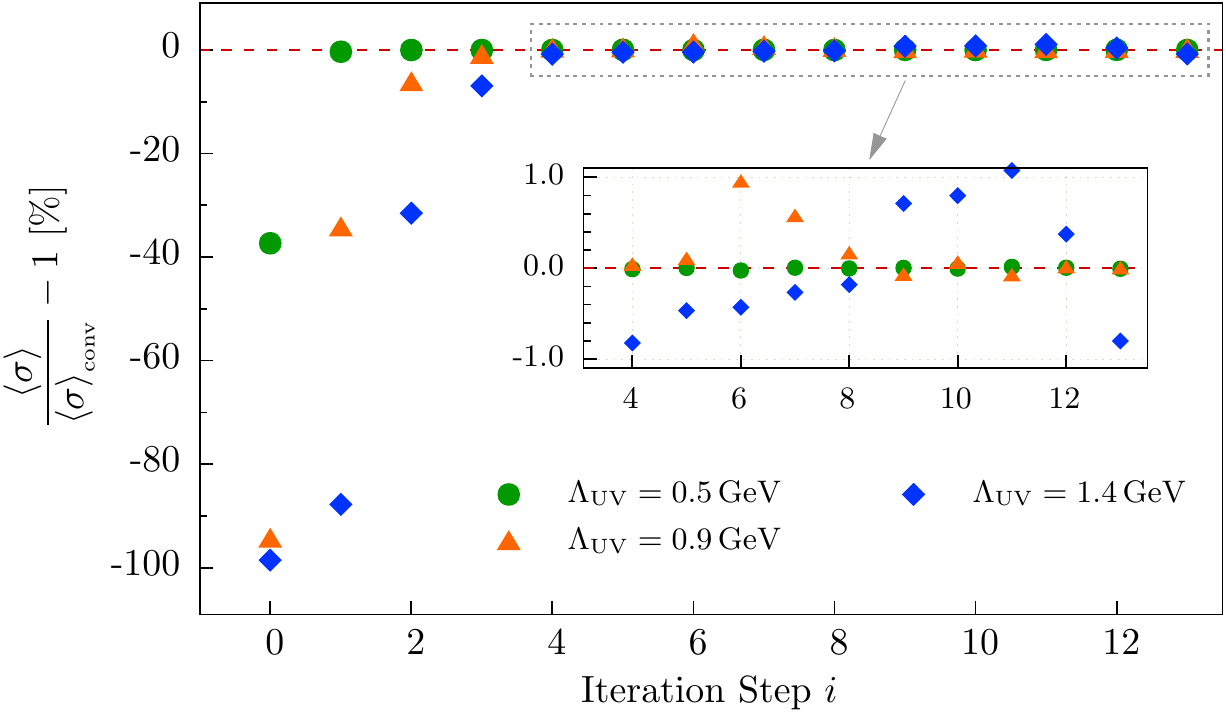}
	\caption[Convergence of the iteration procedure]{Convergence
          of the iteration procedure for different UV cutoff scales.}
	\label{fig:results:convergence_iteration}
\end{figure}

In this appendix, we investigate the iteration procedure's convergence
properties by comparing a given observable obtained from different
iteration steps. Exemplarily, we choose this observable to be the
chiral condensate. Accordingly,
Fig.~\ref{fig:results:convergence_iteration} shows the relative
deviation between the chiral condensate $\chevron{\sigma}$ in step $i$
and the converged result $\chevron{\sigma}_\text{conv}$. We restrict
ourselves to an analysis of the convergence properties at vanishing temperature. 
However, in the finite temperature case the iteration procedure requires
at most one additional iteration step to converge. As pointed out in the main
text keeping the bare or the renormalised Yukawa coupling constant has crucial
impact on the convergence properties of the iteration procedure. As shown in Tab.~\ref{tab:results:massesT0_tune_polbareh} 
in comparison to Tab.~\ref{tab:results:massesT0_tune_polLambda900} in the main text,
the iteration for constant bare Yukawa coupling converges significantly faster than the
corresponding calculation with constant renormalised Yukawa coupling at the same cutoff
scale. Disregarding the fact that a constant bare Yukawa coupling does not correspond to the physical situation,
it is not even possible in this case to find initial conditions for $\Lambda_\text{UV}>\SI{700}{MeV}$ which lead to physical parameters
in the IR due to the stronger running of the wavefunction renormalisation compared to the calculation with fixed renormalised Yukawa coupling.

The dependence of the convergence properties on the UV cutoff scale are illustrated in Fig.~\ref{fig:results:convergence_iteration}. 
If one considers relative deviations of one percent from the full result as approximately converged the calculation at $\Lambda_\text{UV}=\SI{500}{MeV}$ and $\Lambda_\text{UV}=\SI{900}{MeV}$
converge after the first or third iteration step respectively, whereas the the calculation at $\Lambda_\text{UV}= \SI{1.4}{GeV}$ requires four iteration steps to converge.

\begin{table}[hb]
	\centering
	\begin{tabular}{cccc}
          \toprule
          \symhspace{1mm}{step} & \symhspace{1mm}{$m_\text{cur}$ [MeV]} & \symhspace{1mm}{$m_\text{pol}$ [MeV]} & \symhspace{1mm}{$\sigma_\text{min}$ [MeV]}\\
          \midrule
          0 & 202.5 & 202.5 & 93.0\\
          1 & 135.6 & \num{134+-2} & 140.8\\
          2 & 135.5 & \num{135+-2} & 140.8\\
          5 & 135.5 & \num{135+-2} & 140.8\\
          \bottomrule
	\end{tabular}
	\caption{Similar to Tab.~\ref{tab:results:massesT0_tune_polLambda900} for $\Lambda=\SI{900}{MeV}$ but for a fixed bare Yukawa coupling.}
	\label{tab:results:massesT0_tune_polbareh}
\end{table}

\section{Comparison LPA$'$ versus full calculation}
\label{app:lpaprime}

In order to understand why Figs.~\ref{fig:results:sameUVcrossover} and \ref{fig:results:sameIRcrossover} show
only little deviations between the LPA$'$ result and the full calculation with
momentum-dependent meson propagators, we reinvestigate the
corresponding truncations, putting emphasis on the role of the
regulator function within the FRG framework. For the sake of
simplicity, let us thereby start with the situation at vanishing
temperature, where the two-point correlator's momentum-dependent part
can be written as
\begin{equation}
  \Delta\GammaTwo_k(p^2) = Z_k(p^2) p^2 \equiv \bigl( 1+\delta
  Z_k(p^2) \bigr) p^2 \fineq{}
	\label{eq:results:deltaZ}
\end{equation}
due to Euclidean $O(4)$ symmetry. For future reference, we define the
quantity $\delta Z$ via the second equality in
\eqref{eq:results:deltaZ}, which isolates the non-quadratic and hence
non-trivial dependence of the bosonic two-point functions on the
external momentum. Next, recall that one central aspect of the
Wilsonian renormalisation group is the locality of RG flows in
momentum space, which is realised in the FRG via an appropriately
chosen regulator function $R_k$. From a technical point of view, the
calculation of loop integrals therefore effectively involves only loop
momenta $p_\mu$ from within a thin momentum shell around the current
RG scale, \ie $p^2\approx k^2$, as large momentum modes are
regularised by the cutoff insertion $\del[t]R_k$, while fluctuations
with smaller momenta are suppressed by the additional regulator term
in the exact two-point correlator,
$\smash{\GammaTwo_k+R_k}$. Consequently, not the full momentum
dependence of the propagators is resolved in computing RG flows at a
given scale $k$, anyway.

In conclusion, the absence of fully momentum-dependent correlation
functions can be partly compensated for in FRG calculations by
choosing a proper regulator function. Thus, a momentum-independent but
RG scale-dependent wavefunction renormalisation $Z_k$ can be a
reasonable approximation for the full $Z_k(p^2)$.

\begin{figure}[b]
    \centering
    \includegraphics[width=0.97\columnwidth]{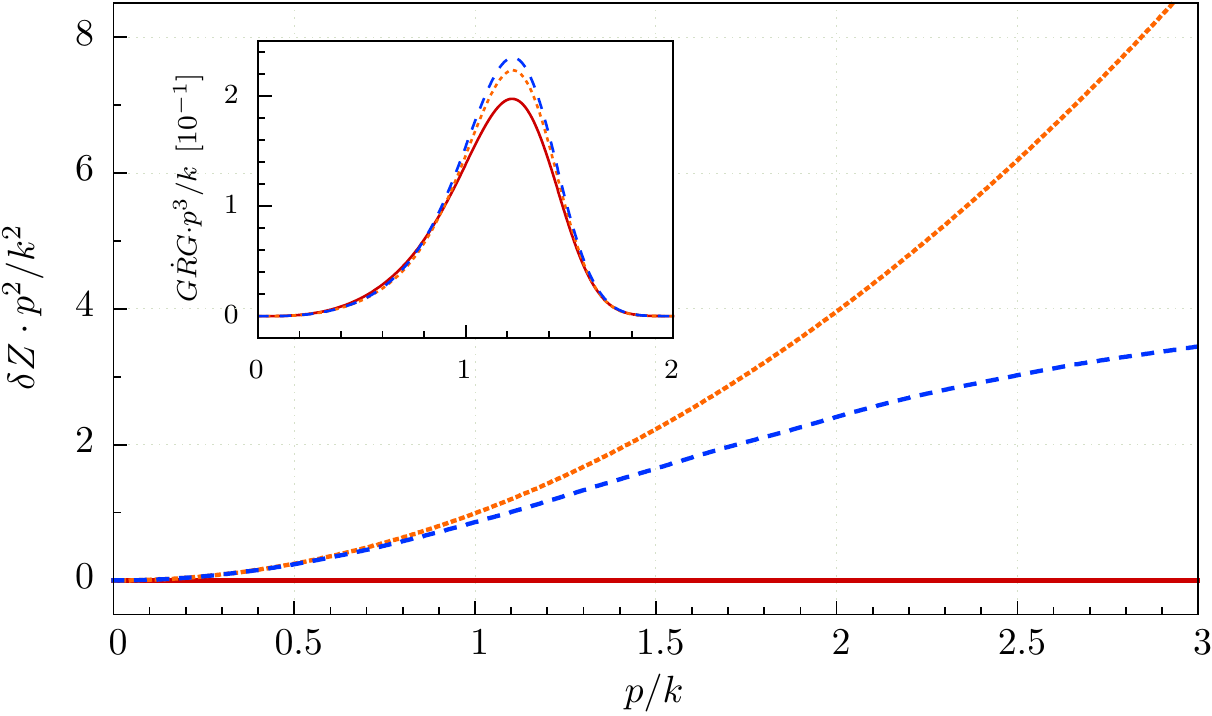}
    \caption[Full momentum dependence]{Momentum dependence of the
      inverse pion propagator. We compare both $\delta Z \cdot p^2$
      and $G\dot{R}G \cdot p^3$ for the full truncation (dashed, blue
      line), LPA$'$ (dotted, orange line) and LPA (solid, red
      line). Exemplarily, we show the case $k=\SI{400}{MeV}$ and $\Lambda_\text{UV} =
      \SI{700}{MeV}$, where the full calculation has a physical IR.}
    \label{fig:results:sameUVmomDepComp_k600}
\end{figure}

The above reasoning can be explicitly verified by studying the
numerically obtained propagators from both LPA$'$ and the full
calculation. In Fig.~\ref{fig:results:sameUVmomDepComp_k600}, we
therefore compare the non-trivial momentum dependencies in the
aforementioned models. In view of the previous discussion, we have
normalised external momenta $\smash{p=\sqrt{p^2}}$ to the given RG
scale $k$. Indeed, one immediately recognises that the correlators
nearly coincide for all $p^2 \lesssim k^2$. Thus, the plot explicitly
demonstrates that some of the propagator's full momentum dependence is
effectively taken into account even by a momentum-independent
approximation such as LPA$'$. Furthermore, deviations between the two
truncations continue to be relatively small in the regime $p/k \in
[1,2]$.

Besides, practical calculations do not involve $\delta Z\cdot p^2$
alone, but rather $Z \cdot p^2$, the dominant behaviour of which is
usually given by the term quadratic in $p^2$, such that the
aforementioned differences will have even less impact on the
outcome. Especially interesting is furthermore the combination $G
\dot{R} G$ appearing \eg in the flow equation of $\del[\rho] U_k$,
which, in turn, is the quantity relevant to calculate the chiral
condensate or the pion curvature mass. We show $(G \dot{R} G)_k(p^2)$
as obtained in the different truncations in the inlay plot of
Fig.~\ref{fig:results:sameUVmomDepComp_k600}.

\newpage

\bibliographystyle{bibstyle}
\bibliography{bib_euclideanit}

\end{document}